\documentclass[10pt,aps,prd,nofootinbib,longbibliography,showpacs]{revtex4-1}

\usepackage{amsmath,bm}
\usepackage{pstricks,pst-node,pst-plot,pst-3d,pst-3dplot,pst-text,pst-eps}
\usepackage{graphicx}

\begin{document}

\title{Many-Body Contributions to Green's Functions and Casimir Energies}
\author{K. V. Shajesh}
\email{shajesh@andromeda.rutgers.edu}
\homepage{http://andromeda.rutgers.edu/~shajesh}
\author{M. Schaden}
\email{mschaden@andromeda.rutgers.edu}
\homepage{http://andromeda.rutgers.edu/~physics/mschaden.htm}
\affiliation{Department of Physics, 
Rutgers, The State University of New Jersey, 
101 Warren Street, Newark, NJ - 07102, USA.}

\date{\today}

\begin{abstract}
The multiple scattering formalism is used to extract irreducible 
$N$-body parts of Green's functions and Casimir energies describing the 
interaction of $N$ objects that are not necessarily mutually disjoint. The 
irreducible $N$-body scattering matrix is expressed in terms of single-body 
transition matrices. The irreducible $N$-body Casimir energy is the trace
of the corresponding irreducible $N$-body part of the Green's function. This
formalism requires the solution of a set of linear integral equations. The
irreducible three-body Green's function and the corresponding Casimir
energy of a massless scalar field interacting with potentials are obtained
and evaluated for three parallel semitransparent plates. When Dirichlet
boundary conditions are imposed on a plate the Green's function and 
Casimir energy decouple into contributions from two disjoint regions.
We also consider weakly interacting triangular--and parabolic-wedges 
placed atop a Dirichlet plate. The irreducible three-body Casimir energy 
of a triangular--and parabolic-wedge is minimal when the shorter side of
the wedge is perpendicular to the Dirichlet plate. The irreducible 
three-body contribution to the vacuum energy is finite and positive in all
the cases studied. 
\end{abstract}

\maketitle

\section{Introduction}

The total energy $E_{12}$ of two static disjoint objects 
may be decomposed as,
\begin{equation}
E_{12} = E_0 + \Delta E_1 + \Delta E_2 + \Delta E_{12},
\label{E12-2}
\end{equation}
where $E_0$ is the energy of the vacuum (medium) without objects,
$\Delta E_1$ and $\Delta E_2$ are (self)-energies required to create 
the objects individually in isolation, and $\Delta E_{12}$ is the change 
in energy due to their interaction.
The interaction energy $\Delta E_{12}$ is finite for 
disjoint objects if it is mediated by an otherwise free quantum field
whose interaction with the objects is described by local potentials.
It is the only contribution to the total energy 
that depends on the position and orientation of \emph{both} objects and 
determines the forces between them. Casimir found that the 
electromagnetic force between two parallel neutral metallic plates does 
not vanish~\cite{Casimir:1948dh} and that the associated Casimir energy 
$\Delta E_{12}$ may be interpreted as arising from changes in the 
zero-point energy due to boundary conditions imposed on the 
electromagnetic field by the metallic plates. Zero-point energy 
contributions to the energies $\Delta E_i$ of 
individual objects in general diverge but the change 
$\Delta E_{12}$ due to the presence of two \emph{disjoint} objects is finite.

Reliable extraction of finite Casimir energies for a long time appeared 
to be restricted to very special geometries,
like parallelepipeds~\cite{Lukosz:1971,Ambjorn:1981xw,Ambjorn:1981xv},
spheres~\cite{Boyer:1968uf,Milton:1978sf}, and 
cylinders~\cite{DeRaad1981229}. 
A multiple scattering formulation for computing Casimir energies of 
smooth objects was developed by 
Balian and Duplantier~\cite{Balian1977300,Balian1978165} 
but it relied heavily on idealized boundary conditions. Kenneth and Klich 
only recently observed~\cite{Kenneth:2006vr} that $\Delta E_{12}$ may 
be computed independent of single-body contributions to the energy 
and is always finite for disjoint objects. This two-body interaction 
energy is compactly expressed~\cite{Kenneth:2006vr,Emig:2007me} in terms of
the free Green's function, $G_0$, and transition operators, $T_1$, $T_2$,
associated with the individual objects,
\begin{equation}
\Delta E_{12} 
= \frac{1}{2} \int_{-\infty}^{\infty}\frac{d\zeta}{2\pi}
\,\text{Tr}\ln\big[ 1- G_0 T_1 G_0 T_2\big]
= \frac{1}{2} \int_{-\infty}^{\infty}\frac{d\zeta}{2\pi}
\,\text{Tr}\ln\big[ 1-\tilde T_1 \tilde T_2\big],
\label{trloggen}
\end{equation}
where we have defined partly amputed transition operators
$\tilde T_i=G_0T_i$, $i=1,2$.
For potential scattering the interaction energy may equivalently be 
expressed~\cite{Milton:2007wz} in terms of the potentials 
$V_i$ and the corresponding Green's functions $G_i$ satisfying 
$G_i V_i=G_0T_i$.

In deriving Eq.~(\ref{trloggen}) one formally subtracts divergent self-energies
and avoids the question of whether these divergences have any physical 
significance. One in particular circumvents the issue raised 
in~\cite{Brown:1969na,Deutsch:1978sc} of how they should be treated 
in the context of gravity, a problem that has so far only been 
considered for parallel 
plates~\cite{Fulling:2007xa,Milton:2007ar,Shajesh:2007sc}.
Although it does not address such conceptual points, 
the irreducible contribution of Eq.~(\ref{trloggen}) suffices to explain 
experimental measurements of Casimir forces between 
two disjoint objects. Since the interaction energy 
for disjoint objects is finite, errors due to numerical or other 
approximations to Eq.~(\ref{trloggen}) can be controlled. This has 
now been demonstrated by explicit calculations for a number of 
geometries and physical situations
\cite{CaveroPelaez:2008tj,CaveroPelaez:2008tk,Maghrebi:2010wp,Maghrebi:2010jx}.

In this article we examine a recently proposed extension of these 
ideas to more than two bodies. It was shown in \cite{Schaden:2010wv} 
that the irreducible $N$-body part of the total energy remains finite 
if the $N$ objects have no {\em common} intersection. We here explicitly 
evaluate the irreducible three-body part, $\Delta E_{123}$, of the total energy,
\begin{equation}
E_{123} = E_0 + \Delta E_1 + \Delta E_2 + \Delta E_3
+ \Delta E_{12} + \Delta E_{23} + \Delta E_{31} + \Delta E_{123},
\label{E-123-ser}
\end{equation}
in several cases and verify that $\Delta E_{123}$ remains finite even as
irreducible two-body contributions diverge. The formal 
expression for $\Delta E_{123}$ given in \cite{Schaden:2010wv} is 
considerably more involved than~Eq.(\ref{trloggen}),
\begin{equation}
\Delta E_{123}
= \frac{1}{2} \int_{-\infty}^{\infty}\frac{d\zeta}{2\pi}
\,\text{Tr}\ln \Big[ 1 + X_{12}
\Big\{ \tilde T_1\tilde T_2\tilde T_3X_{23} 
+\tilde T_1\tilde T_3\tilde T_2X_{32}
-\tilde T_1\tilde T_2\tilde T_1\tilde T_3
-\tilde T_1\tilde T_3\tilde T_2\tilde T_3X_{23}
-\tilde T_1\tilde T_2\tilde T_3\tilde T_2X_{32}
\Big\} X_{13} \Big],
\end{equation}
where the $X_{ij}$'s are solutions to the integral equations,
\begin{equation}
\big[ 1-\tilde T_i\tilde T_j\big]X_{ij}=1.
\label{def-Xij}
\end{equation}

We here obtain and evaluate alternate expressions for 
irreducible Casimir energies that do not involve a logarithm.
The article is organized as follows. In Section~\ref{Nb-G-fun:s} we derive 
Faddeev-like equations for the scattering matrix associated
with $N$-objects and extract their irreducible $N$-body parts.
The associated $N$-body Green's functions are expressed in terms of
one-body transition matrices describing scattering off each 
object individually.  The procedure is illustrated for $N=2$, and $N=3$,
for which explicit solutions are obtained.
The method is generalizable to higher $N$.
The general solutions of Section~\ref{Nb-G-fun:s}
are used to obtain the Green's functions for two and three 
semitransparent plates in Section~\ref{G-fun-semi-Ps:s}.
The irreducible three-body contribution to the Green's function for 
three semitransparent plates is found to exactly cancel the 
two-body interaction of the outer plates when Dirichlet boundary 
conditions are imposed on the central plate.

In Section~\ref{Nb-Cas-En:s} we express the irreducible
$N$-body contribution to the Casimir energy in terms of
the $N$-body part of the transition matrix.
This avoids the computation of the logarithm of an 
integral operator, but requires one to solve a set of linear Faddeev-like
integral equations for the $N$-body transition matrix.
In Section~\ref{Cas-en-semiP:s} we use this formalism
to obtain irreducible two--and three-body contributions to the Casimir energy
of two and three semitransparent plates. 
The three-body contribution again cancels the two-body 
interaction from the outer plates when Dirichlet boundary conditions
are imposed on the central plate.

In Section~\ref{3body-sweak:s}, 
we specialize to the case when two of the three potentials are weak.
For point-potentials we prove that the irreducible two-body Casimir energy
is always negative whereas the irreducible three-body contribution 
is positive. The proof immediately generalizes to any form of the weakly 
interacting potentials. We also derive expressions for irreducible 
two--and three-body contributions to the Casimir energy when 
the weak potentials have translational symmetry and the third potential 
represents a Dirichlet plate parallel to the symmetry axis.
Some of the expressions for the irreducible two-body
contributions appear to not have been noted in earlier studies.
We obtain the irreducible three-body contribution to the Casimir energy
in this semiweak approximation, and verify independently that
it is positive and finite.

In Section~\ref{tri-wedge:s} we use these results
to investigate the irreducible three-body
Casimir energy of a weakly interacting wedge placed atop
a Dirichlet plate forming a waveguide of triangular cross-section.
The potentials forming the triangular waveguide overlap
and the irreducible two-body contributions to the vacuum energy diverge.
However, the irreducible three-body Casimir energy 
is well defined as long as the supports of the three potentials have 
no {\em common} overlap. 
The irreducible three-body Casimir energy is minimal (and vanishes)
when the shorter side of the wedge is perpendicular to the Dirichlet plate.
Inspired by the study in \cite{Abalo:2010ah}, we investigate the 
dependence of the irreducible three-body Casimir energy
on the cross-sectional area and perimeter of the triangular waveguide.
To emphasize that the finiteness of the irreducible three-body Casimir
energies is not only due to the subtraction of corner divergences,
we, in Section~\ref{par-wedge:s}, consider a waveguide with weakly 
interacting sides of parabolic cross-section that touch a Dirichlet plate.
The conclusions for weak triangular-wedges generalize to parabolic-wedges 
with only minor changes in interpretation.

The explicit calculations in this article support the general results of
\cite{Schaden:2010wv}. The irreducible three-body contribution to the 
Casimir energy in the examples considered here is always positive.
It furthermore is continuous (and in this sense is analytic in the 
corresponding parameter) when two of the bodies approach each other
and intersect.


\section{Many-body Green's functions}
\label{Nb-G-fun:s}

The free Green's function of a massless scalar field in Euclidean space-time
satisfies the partial differential equation
\begin{equation}
[-{\bm\nabla}^2 +\zeta^2] G_0({\bf x},{\bf x}^\prime) 
= \delta^{(3)}({\bf x}-{\bf x}^\prime),
\label{G0-def}
\end{equation}
where ${\bm\nabla}^2$ is the Laplacian of flat three-dimensional space.
It is related to the corresponding free Green's function
of Minkowski space-time by a Euclidean (Wick) rotation.
The ``one-body'' Green's function, $G_i$, associated with the 
time-independent potential, $V_i({\bf x})$, 
describing the interaction with the $i$-th object satisfies
\begin{equation}
\left[-{\bm\nabla}^2 +\zeta^2 +V_i({\bf x}) \right]
G_i({\bf x},{\bf x}^\prime) = \delta^{(3)}({\bf x}-{\bf x}^\prime).
\label{Gi-def}
\end{equation}
The two-body Green's function $G_{ij}$ solves a similar equation 
with the potential $(V_i+V_j)$ associated with a pair of objects,
$G_{ijk}$ denotes the three-body Green's function to the potential
$(V_i+V_j+V_k)$, etc.
The potentials $V_i({\bf x})$ of this model~\cite{Bordag:1992cm} 
are proportional to 
$\delta$-functions that simulate the interaction of the scalar field
with classical objects. Infinitely strong $\delta$-function potentials 
enforce Dirichlet boundary conditions at the surface of 
the objects.

One obtains a formal solution to Eq.~(\ref{Gi-def}) by considering the 
differential operator as an integral kernel, and manipulating the 
kernels as if they were matrices. To emphasize the correspondence 
between integral kernels and matrices, we replace:
$[-{\bm\nabla}_{\bf x}^2 +\zeta^2] \,\delta^{(3)}({\bf x}-{\bf x}^\prime)
\to G_0^{-1}$, $V_i({\bf x})\,\delta^{(3)}({\bf x}-{\bf x}^\prime) \to V_i$, 
and $\delta^{(3)}({\bf x}-{\bf x}^\prime) \to 1$.
Using ordinary matrix algebra one obtains the formal solution to 
Eq.~(\ref{Gi-def}) in the form
\begin{equation}
G_i=G_0 - G_0T_iG_0,
\label{Gi-gsol}
\end{equation} 
where the transition matrix $T_i$ is given by 
\begin{equation}
T_i = V_i (1+G_0V_i)^{-1} = (1+V_iG_0)^{-1} V_i
= V_i - V_iG_0V_i + V_iG_0V_iG_0V_i - \ldots.
\label{ti-def}
\end{equation}
The second term in Eq.~(\ref{Gi-gsol}) is interpreted as due to scattering 
off the $i$-th object. It corresponds to the integral operator,
\begin{equation} 
G_0T_iG_0\to{\textstyle\int} d^3x_1 {\textstyle\int} d^3x_1^\prime
\,G_0({\bf x}-{\bf x}_1) T_i({\bf x}_1,{\bf x}_1^\prime)
G_0({\bf x}_1^\prime-{\bf x}^\prime).
\label{G1=G0-RTR}
\end{equation}
In the following, symbolic equations often are more compactly 
written\footnote{This is like setting $G_0=1$.} 
in terms of partly amputated operators.
Equations for the physical operators are obtained by replacing
every partly amputated Greens-function $\tilde G_i$, potential $\tilde V_i$,
and scattering matrix $\tilde T_i$, by,
\begin{equation}
\tilde G_i\to G_iG_0^{-1}, \qquad
\tilde V_i\to G_0V_i, \qquad \text{and} \qquad \tilde T_i\to G_0T_i.
\end{equation}

\subsection{Many-body scattering theory}

The partly amputated $N$-body Green's function satisfies the equation
\begin{equation}
\Big[1 +\tilde V_1 +\tilde V_2 +\ldots+\tilde V_N\Big]\tilde G_{1\ldots N} =1.
\label{Ngeqn-def}
\end{equation}
The numbers in the subscript of $\tilde G_{1\ldots N}$ relate to the
respective potentials.
We may treat the sum of potentials in Eq.~(\ref{Ngeqn-def}) 
as a single potential and thus proceed as for a
single-body. The solution may again be written in the form
\begin{equation}
\tilde G_{1\ldots N}=1-\tilde T_{1\ldots N},
\label{G1-N=M1N}
\end{equation}
where the $N$-body transition matrix $\tilde T_{1\ldots N}$ satisfies the 
equation
\begin{equation}
\Big[ 1 + (\tilde V_1+\tilde V_2+\dots +\tilde V_N)\Big] \tilde T_{1\ldots N}
= (\tilde V_1+\tilde V_2+\dots +\tilde V_N).
\label{T1N-deqn}
\end{equation}
The solution to Eq.~(\ref{T1N-deqn}) is an infinite series similar to 
the one in Eq.~(\ref{ti-def}), whose terms can be regrouped into 
components $\tilde T_{1\ldots N}^{ij}$, that begin with the $i$-th potential
and end with the $j$-th potential. For $N$ potentials we have $N^2$ 
such components representing transitions from the $i$-th to the $j$-th object.
This decomposition of the $N$-body transition matrix is of the form,
\begin{equation}
\tilde T_{1\ldots N} = \sum_{i=1}^N\sum_{j=1}^N \tilde T_{1\ldots N}^{ij}
=\text{Sum} \big[ \tilde{\bf T}_{1\ldots N} \big],
\label{T1N=T1Nij}
\end{equation}
where the symbol $\text{Sum}[{\bf A}]$ stands for the sum of all
elements of the matrix ${\bf A}$.
The matrix form of the $N$-body transition operator is,
\begin{equation}
\tilde{\bf T}_{1\dots N} = \left(
\begin{array}{cccc}
\tilde T_{1\dots N}^{11} & \tilde T_{1\dots N}^{12} 
& \cdots & \tilde T_{1\dots N}^{1N} \\[2mm]
\tilde T_{1\dots N}^{21} & \tilde T_{1\dots N}^{22} 
& \cdots & \tilde T_{1\dots N}^{2N} \\[2mm]
\vdots &\vdots &\ddots & \vdots \\[2mm]
\tilde T_{1\dots N}^{N1} & \tilde T_{1\dots N}^{N2} 
& \cdots & \tilde T_{1\dots N}^{NN}
\end{array} \right),
\end{equation}
where each component is an integral operator.

Inserting Eq.~(\ref{T1N=T1Nij}), and introducing Kronecker-$\delta$ integral
operators, Eq.~(\ref{T1N-deqn}) is equivalent to the following set 
of integral equations
\begin{equation}
\sum_k\Big[ \delta_{ik} +\tilde V_i\Big]\tilde T_{1\ldots N}^{kj}
=\tilde V_i\delta_{ij}.
\end{equation}
In matrix notation this set of equations is 
\begin{equation}
\big[ {\bf 1}+ \tilde{\bf V}_\text{diag} + \tilde{\bf\Theta}_{1\ldots N}^V\big]
\cdot \tilde{\bf T}_{1\ldots N} = \tilde{\bf V}_\text{diag},
\label{1vtt=v}
\end{equation}
where we have introduced general matrix symbols
\begin{equation}
{\bm \Theta}_{1\ldots N}^A = \left(
\begin{array}{ccccc}
0 & A_1 & A_1 & \cdots & A_1 \\[1mm]
A_2 & 0 & A_2 & \cdots & A_2 \\[1mm]
A_3 & A_3 & 0 & \cdots & A_3 \\[1mm]
\vdots &\vdots &\vdots &\ddots & \vdots \\[1mm]
A_N & A_N & A_N & \cdots & 0
\end{array} \right),
\qquad \qquad
{\bm A}_\text{diag} = \left(
\begin{array}{ccccc}
A_1 & 0 & 0 & \cdots & 0 \\[1mm]
0 & A_2 & 0 & \cdots & 0 \\[1mm]
0 & 0 & A_3 & \cdots & 0 \\[1mm]
\vdots &\vdots &\vdots &\ddots & \vdots \\[1mm]
0 & 0 & 0 & \cdots & A_N 
\end{array} \right).
\end{equation}

Using these definitions with Eq.~(\ref{ti-def}) we write,
\begin{equation}
\big[{\bf 1}+\tilde{\bf V}_\text{diag} \big]
\cdot \tilde{\bf T}_\text{diag} = \tilde{\bf V}_\text{diag}
\qquad \text{and} \qquad
\big[ {\bf 1} + \tilde{\bf V}_\text{diag} \big] 
\cdot \tilde{\bm \Theta}_{1\ldots N}^T =\tilde{\bf\Theta}_{1\ldots N}^V,
\end{equation}
and use in Eq.~(\ref{1vtt=v}) to derive 
\begin{equation}
\big[{\bf 1} +\tilde{\bm \Theta}_{1\ldots N}^T \big]
\cdot\tilde{\bf T}_{1\dots N}= \tilde{\bf T}_\text{diag}.
\label{fadeqn}
\end{equation}
The set of linear integral equations of Eq.~(\ref{fadeqn}) are often
referred to as Faddeev's equations~\cite{Faddeev:1965a, Merkuriev:1993a}
for nuclear many-body scattering, but apparently have been 
known~\cite{francis:1953a, brueckner:1954a}
in the context of ``optical models'' for atomic nuclei since the 1950's
and may have been used in earlier optical studies. 
The closely related approach in \cite{Martin:1959a, Puff1961317} 
is known as Martin-Schwinger-Puff many-body theory.

Eq.~(\ref{G1-N=M1N}) relates the $N$-body Green's function
$\tilde G_{1,\dots N}$ to the $N$-body transition matrix 
$\tilde T_{1,\dots N}$ satisfying Eq.~(\ref{T1N-deqn}).
Faddeev's equations of Eq.~(\ref{fadeqn}) reduce the problem 
of solving Eq.~(\ref{T1N-deqn}) for the $N$-body transition matrix
to that of inverting 
$\big[{\bf 1}+\tilde{\bf \Theta}_{1\ldots N}^T \big]$
by solving a set of $N$ linear integral equations.
Remarkably, ${\bf \Theta}_{1\ldots N}^T$ depends only on
single-body transition operators. 
The norm of ${\bf \Theta}_{1\ldots N}^T$ is less than unity 
(because the norm of single-body transition matrices is) 
and Faddeev's equations can, at least in principle, be solved by 
(numerical) iteration~\cite{Merkuriev:1993a}.

\subsection{$N=2$: Two-body interaction}

Using Eq.~(\ref{Ngeqn-def}) the Green's function equation for $N=2$
has solution given by Eq.~(\ref{G1-N=M1N}), where
the transition matrix is obtained by inverting the Faddeev's 
equation in Eq.~(\ref{fadeqn}) to yield
\begin{equation}
\tilde{\bf T}_{12} = \big[ {\bf 1}+ \tilde{\bf \Theta}_{12}^T\big]^{-1} 
\cdot \tilde{\bf T}_\text{diag}
= \left[ \begin{array}{cc} X_{12} & 0 \\ 0 & X_{21} \end{array} \right]
\left[ \begin{array}{cc} \tilde T_1 & -\tilde T_1\tilde T_2 \\
-\tilde T_2\tilde T_1 & \tilde T_2 \end{array}\right].
\label{12-T12}
\end{equation}
The integral operators $X_{ij}$ in Eq.~(\ref{12-T12})
satisfy Eq.~(\ref{def-Xij}).
Summing the components of $\tilde{\bf T}_{12}$
we obtain the total transition matrix as,
\begin{equation}
\tilde T_{12} = \text{Sum}\big[ \tilde{\bf T}_{12} \big]
= \big[ 1-X_{12}\tilde G_1\big] + \big[ 1-X_{21}\tilde G_2\big].
\end{equation}

The total two-body transition matrix $\tilde {\bf T}_{12}$ can be 
decomposed into its irreducible one--and two-body parts,
\begin{equation}
\tilde {\bf T}_{12} = \tilde {\bf T}_1 + \tilde {\bf T}_2 
+ \Delta \tilde {\bf T}_{12},
\label{T12=T1T2DT12}
\end{equation}
with
\begin{equation}
\tilde {\bf T}_{1} 
= \left[ \begin{array}{cc} \tilde T_1 & 0 \\ 0 & 0 \end{array}\right],
\qquad
\tilde {\bf T}_{2} 
= \left[ \begin{array}{cc} 0 & 0 \\ 0 & \tilde T_2 \end{array}\right],
\qquad
\Delta \tilde{\bf T}_{12} =
\left[ \begin{array}{cc} X_{12} & 0 \\ 0 & X_{21} \end{array} \right]
\left[ \begin{array}{cc} 
\tilde T_1\tilde T_2\tilde T_1 & -\tilde T_1\tilde T_2 \\
-\tilde T_2\tilde T_1 & \tilde T_2\tilde T_1\tilde T_2
\end{array}\right].
\label{DT12=expl}
\end{equation}
The irreducible two-body transition matrix $\Delta \tilde{\bf T}_{12}$
includes all contribution with scattering off \emph{both} potentials.
Eq.~(\ref{T12=T1T2DT12}) and the definition in Eq.~(\ref{G1-N=M1N})
imply the following decomposition of the partly amputated 
two-body Green's function
\begin{equation}
\tilde G_{12}=1-\tilde T_1-\tilde T_2-\Delta \tilde T_{12},
\label{G12-casc}
\end{equation}
where $\Delta \tilde T_{12}=\text{Sum}\big[\Delta \tilde{\bf T}_{12}\big]$.
Summing the four independent two-body transitions of Eq.~(\ref{DT12=expl}),
the irreducible two-body transition operator in terms of the 
single-body transition matrices $\tilde T_1$ and $\tilde T_2$ is,
\begin{equation}
\Delta \tilde T_{12} = \text{Sum}\big[\Delta \tilde{\bf T}_{12}\big] 
= (1-X_{12})\tilde G_1 + (1-X_{21})\tilde G_2.
\label{DT12full}
\end{equation}

\subsection{$N=3$: Three-body interaction}

One proceeds similarly for three bodies. In this case the formal solution
to the Faddeev's equation, of Eq.~(\ref{fadeqn}), is
\begin{equation}
\tilde{\bf T}_{123} =
\left[ \begin{array}{ccc}
X_{1[23]} & 0 & 0 \\[1mm] 0 & X_{2[31]} & 0 \\[1mm] 0 & 0 & X_{3[12]} 
\end{array}\right]
\left[ \begin{array}{ccc}
\tilde T_1 & -\tilde T_1\tilde G_3\tilde T_2X_{32} 
& -\tilde T_1\tilde G_2\tilde T_3X_{23} \\[1mm]
-\tilde T_2\tilde G_3\tilde T_1X_{31} & \tilde T_2 
& -\tilde T_2\tilde G_1\tilde T_3X_{13} \\[1mm]
-\tilde T_3\tilde G_2\tilde T_1X_{21} 
& -\tilde T_3\tilde G_1\tilde T_2X_{12} & \tilde T_3
\end{array}\right],
\label{123-rtr}
\end{equation}
where the $\tilde G_i$'s are related to $\tilde T_i$'s by Eq.~(\ref{Gi-gsol})
and the two-body effective Green's functions, $X_{ij}$, ($i\neq j$),
solve Eq.~(\ref{def-Xij}).
The three-body effective Green's functions, $X_{i[jk]}$, 
($i\neq j\neq k$), satisfy the equation
\begin{equation}
X_{i[jk]} \big[ 1 -\tilde T_i\tilde T_{jk} \big]
=X_{i[jk]} \big[ 1 -\tilde T_i\tilde G_j\tilde T_k X_{jk}
-\tilde T_i\tilde G_k\tilde T_j X_{kj} \big] = 1.
\label{Xijk-def}
\end{equation}
The total transition matrix in this case is
\begin{equation}
\tilde T_{123} = \text{Sum}\big[ \tilde{\bf T}_{123} \big]
= \big[ 1 - X_{1[23]}\tilde G_1\big]
+ \big[ 1 - X_{2[31]}\tilde G_2\big] + \big[ 1 - X_{3[12]}\tilde G_3\big].
\end{equation}

The transition matrix may again be decomposed into irreducible parts
\begin{equation}
\tilde{\bf T}_{123} = \tilde{\bf T}_1 + \tilde{\bf T}_2 + \tilde{\bf T}_3
+ \Delta \tilde{\bf T}_{12}+\Delta\tilde{\bf T}_{23}+\Delta\tilde{\bf T}_{31}
+ \Delta \tilde{\bf T}_{123},
\label{T123=T1..DT123}
\end{equation}
where 
\begin{equation}
\Delta {\bf\tilde T}_{12}+\Delta {\bf\tilde T}_{23}+\Delta {\bf\tilde T}_{31}
=\left[ \begin{array}{ccc}
X_{12}\tilde T_1\tilde T_2\tilde T_1 
+X_{13}\tilde T_1\tilde T_3\tilde T_1& 
-X_{12}\tilde T_1\tilde T_2 &-X_{13}\tilde T_1\tilde T_3 \\[1mm]
-X_{21}\tilde T_2\tilde T_1 & X_{21}\tilde T_2\tilde T_1\tilde T_2
+X_{23}\tilde T_2\tilde T_3\tilde T_2 &-X_{23}\tilde T_2\tilde T_3 \\[1mm]
-X_{31}\tilde T_3\tilde T_1 
&-X_{32}\tilde T_3\tilde T_2 & X_{31}\tilde T_3\tilde T_1\tilde T_3
+X_{32}\tilde T_3\tilde T_2\tilde T_3
\end{array}\right]
\end{equation}
is obtained using the $N=2$ expressions of Eq.~(\ref{DT12=expl}).
The new irreducible three-body part is,
\begin{equation}
\Delta \tilde{\bf T}_{123} =
\left[ \begin{array}{ccc}
(1 -X_{12} -X_{13}+X_{1[23]})\tilde T_1 
 & [\tilde T_1X_{21} -X_{1[23]}\tilde T_1\tilde G_3X_{23}]\tilde T_2
 & [\tilde T_1X_{31} -X_{1[23]}\tilde T_1\tilde G_2X_{32}]\tilde T_3 \\[1mm]
[\tilde T_2X_{12} -X_{2[31]}\tilde T_2\tilde G_3X_{13}]\tilde T_1
  & (1-X_{23}-X_{21}+X_{2[31]})\tilde T_2
 & [\tilde T_2X_{32} -X_{2[31]}\tilde T_2\tilde G_1X_{31}]\tilde T_3 \\[1mm]
[\tilde T_3X_{13} -X_{3[12]}\tilde T_3\tilde G_2X_{12}]\tilde T_1
  & [\tilde T_3X_{23} -X_{3[12]}\tilde T_3\tilde G_1X_{21}]\tilde T_2
  & (1-X_{31}-X_{32}+X_{3[12]})\tilde T_3
\end{array}\right].
\label{d123-rtr}
\end{equation}
The decomposition of Eq.~(\ref{T123=T1..DT123}) carries over to 
the decomposition of the Green's function
\begin{equation}
\tilde G_{123} 
=1 - \tilde T_1 - \tilde T_2 - \tilde T_3 
- \Delta \tilde T_{12} - \Delta \tilde T_{23} - \Delta \tilde T_{31} 
- \Delta \tilde T_{123}.
\label{G123-casc}
\end{equation}
Summing the nine independent three-body transitions in 
Eq.~(\ref{d123-rtr}) we find that, 
\begin{equation}
\Delta \tilde T_{123} = \text{Sum}\big[\Delta \tilde{\bf T}_{123}\big] 
= -(1-X_{12}-X_{13}+X_{1[23]})\tilde G_1 
 -(1-X_{23}-X_{21}+X_{2[31]})\tilde G_2 
 -(1-X_{31}-X_{32}+X_{3[12]})\tilde G_3. 
\end{equation}
Although not quite as obvious as for two-body scattering,
closer inspection reveals that each component of $\Delta \tilde{\bf T}_{123}$
indeed involves scattering off all three bodies. A similar procedure
can be used to obtain scattering matrices and their irreducible parts
for more than three bodies.

\section{Green's functions for parallel semitransparent $\delta$-plates}
\label{G-fun-semi-Ps:s}

We now apply this formalism to derive the Green's functions 
for parallel semitransparent plates of infinite extent described by 
$\delta$-function potentials
\begin{equation}
V_i({\bf x}) = \lambda_i \delta(z-a_i),
\label{Vi-def}
\end{equation}
where $a_i$ specifies the position of the $i$-th plate on the $z$-axis, and
$\lambda_i>0$ is the coupling parameter. In the limit 
$\lambda_i\to\infty$ the potential of Eq.~(\ref{Vi-def})
simulates a plate with Dirichlet boundary conditions.
The translation symmetry in the $x$-$y$ plane can be exploited and
Eq.~(\ref{Gi-def}) written in terms of the dimensionally reduced 
Green's function, $g_i(z,z^\prime)$, defined by 
\begin{equation}
G_i({\bf x},{\bf x}^\prime;\zeta) = \int\frac{d^2k}{(2\pi)^2}
\,e^{i{\bf k}_\perp\cdot({\bf x}-{\bf x}^\prime)_\perp} g_i(z,z^\prime;\kappa),
\label{Gi=Igi}
\end{equation}
where ${\bf x}_\perp$ is the component of ${\bf x}$ in the
$x$-$y$ plane. ${\bf k}_\perp$ is the corresponding Fourier component,
and $\kappa^2=\zeta^2+{\bf k}^2_\perp$, ${\bf k}_\perp^2=k_x^2+k_y^2$.
Since the potentials of Eq.~(\ref{Vi-def}) do not depend on the 
transverse dimensions, the Green's functions $G_{1\dots N}$ 
for $N$ parallel plates also correspond to dimensionally 
reduced $g_{1\dots N}$.

\subsection{$N=1$: Green's function for a single semitransparent plate}

When substituted in Eq.~(\ref{Gi-def}),  Eq.~(\ref{Gi=Igi})
implies that $g_i(z,z^\prime)$ solves a one-dimensional ordinary
inhomogeneous second order differential equation with a $\delta$-function
potential that can be solved explicitly, to obtain,
\begin{equation}
g_i(z,z^\prime) = \frac{1}{2\kappa}\,e^{-\kappa|z-z^\prime|}
-\frac{\bar{\lambda}_i}{1+\bar{\lambda}_i}
\frac{1}{(2\kappa)^2}
\,e^{-\kappa|z-a_i|}e^{-\kappa|z^\prime-a_i|},
\label{g1-exsol}
\end{equation}
where $\bar{\lambda}_i=\lambda_i/2\kappa$,
and $\kappa$ was defined after Eq.~(\ref{Gi=Igi}).
We also arrive at this solution by starting from Eq.~(\ref{Gi-gsol}),
which for the dimensionally reduced Green's function reads
\begin{equation}
g_i(z,z^\prime) = g_0(z-z^\prime) - r_i(z) t_i r_i(z^\prime),
\label{g1=g0t1}
\end{equation}
where $g_0(z,z^\prime)$ is the dimensionally reduced free Green's function,
and $r_i(z)=g_0(z-a_i)$. 
Eq.~(\ref{G0-def}) implies that
\begin{equation}
g_0(z,z^\prime) = \frac{1}{2\kappa}\,e^{-\kappa|z-z^\prime|},
\quad r_i(z) = \frac{1}{2\kappa}\,e^{-\kappa|z-a_i|}.
\end{equation}
It will be convenient to define,
\begin{equation}
\bar{r}_i(z)=2\kappa\, r_i(z)=e^{-\kappa |z-a_i|}, 
\quad \text{and} \quad
\bar{r}_{ij}=e^{-\kappa |a_i-a_j|}=e^{-\kappa a_{ij}},
\end{equation}
where for notational convenience we have defined $a_{ij}=|a_i-a_j|$.
The dimensionally reduced transition matrix in Eq.~(\ref{g1=g0t1})
is found by summing the series in Eq.~(\ref{ti-def}).
Translational invariance in transverse directions and the 
$\delta$-function potential render all integrals trivial and the 
series can be re-summed to give
\begin{equation}
t_i(z,z^\prime) = 2\kappa\, \bar{t}_i
\,\delta(z-a_i)\delta(z^\prime-a_i), 
\qquad \bar{t}_i= \frac{\bar{\lambda}_i}{1+\bar{\lambda}_i}.
\label{g1-exp}
\end{equation}
In the Dirichlet limit ($\lambda_i\to\infty$) the transition matrix
simplifies further to $\bar{t}_i^D=1$. Inserting the dimensionally 
reduced transition matrix of Eq.~(\ref{g1-exp}) in Eq.~(\ref{g1=g0t1})
reproduces the explicit solution of Eq.~(\ref{g1-exsol}).

\subsection{$N=2$: 
Green's function for parallel semitransparent plates}

The previous procedure is readily extended to compute the Green's
function of $N$ semitransparent plates located at 
$z=a_i$, $i=1,2,\ldots,N$, and described by potentials of the form
given by Eq.~(\ref{Vi-def}) with associated strengths $\lambda_i$.
Generalization of Eq.~(\ref{g1=g0t1}) in particular gives the relation
\begin{equation}
g_{1\ldots N}(z,z^\prime) = g_0(z-z^\prime) 
- {\bf r}(z)^T \cdot {\bf t}_{1\ldots N} \cdot {\bf r}(z^\prime),
\label{12-rtr}
\end{equation}
between the dimensionally reduced Green's function $g_{1\ldots N}(z,z^\prime)$
and the corresponding components of the dimensionally reduced 
transition matrix. The vector ${\bf r}(z)$ constructed out of the 
dimensionally reduced free Green's function originating or ending 
at a plate is given by
\begin{equation}
{\bf r}(z)^T = \left[ r_1(z), r_2(z),\, \ldots\,, r_N(z) \right]
= \frac{1}{2\kappa} \left[
e^{-\kappa |z-a_1|}, e^{-\kappa |z-a_2|},\,\ldots\,,e^{-\kappa |z-a_N|}\right].
\label{vecrz}
\end{equation}
An advantage of this approach is that the Faddeev integral equations,
Eq.~(\ref{fadeqn}), collapse to algebraic equations for the dimensionally
reduced transition matrix ${\bf t}_{1\ldots N}$ due to the translational 
symmetry and the $\delta$-function potentials. 
The transition matrix decouples from the ${\bf r}$-vector, which leads 
to considerable simplification in the evaluation of the Green's function. A 
similar simplification occurs for concentric cylinders and concentric spheres.

The dimensionally reduced two-body transition matrix can be read out from 
Eq.~(\ref{12-T12}) once the corresponding $X_{ij}$ has been evaluated.
With the single-body transition matrices of Eq.~(\ref{g1-exp}) all integrals
evaluate trivially and the solution of Eq.~(\ref{def-Xij}) for $X_{ij}$ is
\begin{equation}
X_{ij}=X_{ji}=\frac{1}{\Delta_{ij}}, 
\qquad \Delta_{ij} = 1-\bar t_i\bar r_{ij}\bar t_j\bar r_{ji} 
= 1 - \bar{t}_i\bar{t}_j\,e^{-2\kappa a_{ij}}.
\label{Xij-sol}
\end{equation}
Using Eq.~(\ref{12-T12}), the dimensionally reduced transition matrix 
for two plates is, 
\begin{equation}
{\bf t}_{ij} = \frac{2\kappa}{\Delta_{ij}}
\left[ \begin{array}{cc}
\bar{t}_i & -\bar{t}_i\bar{r}_{ij}\bar{t}_j \\
-\bar{t}_j\bar{r}_{ji}\bar{t}_i & \bar{t}_j
\end{array}\right].
\label{t12i2p}
\end{equation}
Eq.~(\ref{t12i2p}) inserted in Eq.~(\ref{12-rtr})
gives the Green's function for two semitransparent parallel plates.
In the Dirichlet limit, $\bar{t}_i\to\bar{t}_i^D=1$,
we have $\Delta_{ij}\to\Delta_{ij}^D= (1-e^{-2\kappa a_{ij}})$,
and the transition matrix for two Dirichlet plates simplifies to, 
\begin{equation}
{\bf t}_{ij}^D = \frac{\kappa}{\sinh\kappa a_{ij}}
\left[ \begin{array}{cc} e^{\kappa a_{ij}} & -1 \\ -1 & e^{\kappa a_{ij}} 
\end{array}\right].
\end{equation}
From Eq.~(\ref{T12=T1T2DT12}) we similarly obtain the irreducible two-body
part of the dimensionally reduced transition matrix as 
\begin{equation}
\Delta {\bf t}_{ij} = -2\kappa \left[ \begin{array}{cc}
\bar{t}_i \left\{1-\frac{1}{\Delta_{ij}}\right\} \hspace{5mm}
& \bar{t}_i\frac{\bar{r}_{ij}}{\Delta_{ij}}\bar{t}_j \\[2mm]
\bar{t}_j\frac{\bar{r}_{ji}}{\Delta_{ji}}\bar{t}_i 
& \bar{t}_j \left\{1-\frac{1}{\Delta_{ij}}\right\} 
\end{array}\right]
= \frac{2\kappa}{\Delta_{ij}} \left[ \begin{array}{cc}
\bar{t}_i \bar{r}_{ij} \bar{t}_j \bar{r}_{ji} \bar{t}_i 
& -\bar{t}_i\bar{r}_{ij}\bar{t}_j \\[2mm]
-\bar{t}_j\bar{r}_{ji}\bar{t}_i 
& \bar{t}_j \bar{r}_{ji} \bar{t}_i \bar{r}_{ij} \bar{t}_j 
\end{array}\right],
\label{Dt12=sol}
\end{equation}
which in the Dirichlet limit simplifies to 
\begin{equation}
\Delta {\bf t}_{ij}^D = \frac{\kappa}{\sinh\kappa a_{ij}}
\left[ \begin{array}{cc} e^{-\kappa a_{ij}} & -1 \\ -1 & e^{-\kappa a_{ij}}
\end{array}\right].
\label{t12D=exp}
\end{equation}

The two-plate Green's function has been obtained 
previously~\cite{CaveroPelaez:2008tj} in a more direct manner.
We reproduced it using the multiple scattering method because 
this approach readily generalizes to more than two plates.

\subsection{$N=3$: 
Green's function for three parallel semitransparent plates}

\begin{center}
\begin{figure}
\includegraphics[scale=1.0]{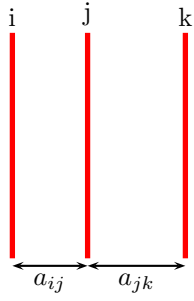}
\caption{Three parallel plates. Plates $i$ and $k$ are separated by
distances $a_{ij}$ and $a_{jk}$ from the center plate $j$.}
\label{3paraPl}
\end{figure}
\end{center}

The three semitransparent plates $i$, $j$, and $k$, of infinite extent
and parallel to the $xy$-plane are described by potentials of the form
given in Eq.~(\ref{Vi-def}). Without loss of generality we assume that
$a_i<a_j<a_k$ (see FIG. \ref{3paraPl}). In the previously introduced
notation this implies that $a_{ij}+a_{jk}=a_{ik}$.
The vector ${\bf r}(z)$ in Eq.~(\ref{vecrz}) now has three components.
 
The dimensionally reduced three-body transition matrix is obtained by
solving Eq.~(\ref{Xijk-def}) for $X_{i[jk]}$ using the $X_{ij}$
of Eq.~(\ref{Xij-sol}). For three semitransparent parallel plates 
one finds that,
\begin{equation}
X_{i[jk]}X_{jk} = \frac{1}{\Delta_{ijk}},
\qquad
\Delta_{ijk} = 1 -\bar{t}_i\bar{r}_{ij}\bar{t}_j\bar{r}_{ji} 
-\bar{t}_j\bar{r}_{jk}\bar{t}_k\bar{r}_{kj} 
-\bar{t}_k\bar{r}_{ki}\bar{t}_i\bar{r}_{ik} 
+2\bar{t}_i\bar{r}_{ij}\bar{t}_j\bar{r}_{jk}\bar{t}_k\bar{r}_{ki}.
\label{Del123-def}
\end{equation}
Using Eq.~(\ref{Del123-def}) in Eq.~(\ref{123-rtr}) we obtain 
\begin{equation}
{\bf t}_{ijk} = \frac{2\kappa}{\Delta_{ijk}}
\left[ \begin{array}{ccc}
\bar{t}_i\Delta_{jk}
&-\bar{t}_i\bar r_{ij[k]}\bar{t}_j &-\bar{t}_i\bar r_{ik[j]}\bar{t}_k \\[2mm]
-\bar{t}_j\bar r_{ji[k]}\bar{t}_i & \bar{t}_j\Delta_{ki}
&-\bar{t}_j\bar r_{jk[i]}\bar{t}_k \\[2mm]
-\bar{t}_k\bar r_{ki[j]}\bar{t}_i &-\bar{t}_k\bar r_{kj[i]}\bar{t}_j 
&\bar{t}_k\Delta_{ij}
\end{array}\right],
\label{t123=expr}
\end{equation}
where 
\begin{equation}
\bar r_{ij[k]} =\bar r_{ij}-\bar r_{ik} \bar{t}_k \bar r_{kj}.
\end{equation}
It is apparent from the solutions for the $N=2$ and $N=3$ case that
terms contributing to the transition matrix in the multiple scattering
expansion depend exponentially on the length of the path of propagation.
Expanding the inverse determinants $\Delta_{ij}$ and $\Delta_{ijk}$
gives all paths contributing to the scattering operator.
This is particularly transparent in our essentially 1-dimensional example.

From Eq.~(\ref{d123-rtr}) the dimensionally reduced, irreducible,
three-body transition matrix is similarly evaluated as,
\begin{equation}
\Delta {\bf t}_{ijk} = 2\kappa \left[ \begin{array}{ccc}
\bar{t}_i \left\{1 -\frac{1}{\Delta_{ij}} -\frac{1}{\Delta_{ik}}
+\frac{\Delta_{jk}}{\Delta_{ijk}} \right\} 
& \bar{t}_i \left\{ \frac{\bar r_{ij}}{\Delta_{ij}} 
-\frac{\bar r_{ij[k]}}{\Delta_{ijk}} \right\} \bar{t}_j
& \bar{t}_i \left\{ \frac{\bar r_{ik}}{\Delta_{ik}} 
-\frac{\bar r_{ik[j]}}{\Delta_{ijk}} \right\} \bar{t}_k
\\[3mm]
\bar{t}_j \left\{ \frac{\bar r_{ji}}{\Delta_{ji}} 
-\frac{\bar r_{ji[k]}}{\Delta_{ijk}} \right\} \bar{t}_i
&\bar{t}_j \left\{1 -\frac{1}{\Delta_{ji}} -\frac{1}{\Delta_{jk}}
+\frac{\Delta_{ik}}{\Delta_{ijk}} \right\} 
& \bar{t}_j \left\{ \frac{\bar r_{jk}}{\Delta_{jk}} 
-\frac{\bar r_{jk[i]}}{\Delta_{ijk}} \right\} \bar{t}_k
\\[3mm]
\bar{t}_k \left\{ \frac{\bar r_{ki}}{\Delta_{ki}} 
-\frac{\bar r_{ki[j]}}{\Delta_{ijk}} \right\} \bar{t}_i
& \bar{t}_k \left\{ \frac{\bar r_{kj}}{\Delta_{kj}} 
-\frac{\bar r_{kj[i]}}{\Delta_{ijk}} \right\} \bar{t}_j
& \bar{t}_k \left\{1 -\frac{1}{\Delta_{ki}} -\frac{1}{\Delta_{kj}}
+\frac{\Delta_{ij}}{\Delta_{ijk}} \right\}
\end{array}\right].
\label{Delt123-3psol}
\end{equation}

It is interesting to consider the situation when Dirichlet boundary 
conditions hold on the $j$-th plate {\em between} the other two plates.
Taking the limit $\bar{t}_j\to \bar{t}_j^D=1$, the
determinant for three parallel plates is found to factorize 
into a product of two-body determinants,
\begin{equation}
\Delta_{i\infty k}= \Delta_{i\infty} \Delta_{k\infty}
=(1-\bar{t}_i\,e^{-2\kappa a_{ij}}) (1-\bar{t}_k\,e^{-2\kappa a_{jk}}),
\label{det-fact}
\end{equation}
where replacing the subscript of a plate by $\infty$ denotes 
Dirichlet boundary conditions on that plate. In this situation,
$\bar r_{ik[\infty]}=(1-\bar{t}_j^D)\bar r_{ik}=0$,
$\bar r_{i\infty [k]}=e^{\kappa a_{ij}}\Delta_{k\infty}$,
$\bar r_{k\infty [i]}=e^{\kappa a_{jk}}\Delta_{i\infty}$,
and Eq.~(\ref{t123=expr}) simplifies to 
\begin{equation}
{\bf t}_{i\infty k} = 2\kappa
\left[ \begin{array}{ccc}
\bar{t}_i \frac{1}{\Delta_{i\infty}}
&-\bar{t}_i \frac{\bar r_{ij}}{\Delta_{i\infty}} &0\\[2mm]
-\frac{\bar r_{ji}}{\Delta_{\infty i}} \bar{t}_i \hspace{2mm}
& \frac{1}{\Delta_{i\infty}} + \frac{1}{\Delta_{k\infty}} -1 \hspace{2mm}
&-\frac{\bar r_{jk}}{\Delta_{k\infty}} \bar{t}_k \\[2mm]
0&-\bar{t}_k \frac{\bar r_{kj}}{\Delta_{k\infty}}
& \bar{t}_k \frac{1}{\Delta_{k\infty}}
\end{array}\right].
\end{equation}
This leads to the observation that
\begin{equation}
{\bf t}_{i\infty k}
={\bf t}_{i\infty} +{\bf t}_{k\infty}-{\bf t}_j^D
= {\bf t}_i + {\bf t}_j^D + {\bf t}_k
+ \Delta {\bf t}_{i\infty} + \Delta {\bf t}_{k\infty}.
\label{tiDk=allts}
\end{equation}
Comparing Eq.~(\ref{tiDk=allts}) 
with the decomposition of the three-body transition matrix
into irreducible one--and two-body parts in Eq.~(\ref{T123=T1..DT123})
this implies
\begin{equation}
\Delta {\bf t}_{i\infty k} +\Delta {\bf t}_{ik} = 0,
\label{etijk=2tik}
\end{equation}
which confirms the notion that modes in the two half-spaces on either side 
of a Dirichlet plate are independent and that correlations between them
must vanish. The irreducible three-body correlations in this limit therefore 
must cancel irreducible two-body correlations between objects on 
opposite sides of the plate. Taking the Dirichlet limit on the central plate 
in Eq.~(\ref{Delt123-3psol}) this is verified explicitly,
\begin{equation}
\Delta {\bf t}_{i\infty k} = 2\kappa \left[ \begin{array}{ccc}
\bar{t}_i \left\{1 -\frac{1}{\Delta_{ik}} \right\} & 0
& \bar{t}_i \frac{\bar r_{ik}}{\Delta_{ik}} \bar{t}_k \\[3mm]
0 & 0 & 0 \\[3mm]
\bar{t}_k \frac{\bar r_{ki}}{\Delta_{ki}} \bar{t}_i & 0
& \bar{t}_k \left\{1 -\frac{1}{\Delta_{ki}} \right\}
\end{array}\right] = -\Delta {\bf t}_{ik},
\label{tiDk=-tik}
\end{equation}
where we have used Eq.~(\ref{Dt12=sol}).

Let us finally consider the case when Dirichlet boundary conditions
are imposed on all three plates. The three-body determinant again
factorizes, 
$\Delta_{ijk}^D = \Delta_{ij}^D \Delta_{jk}^D 
=(1-e^{-2\kappa a_{ij}}) (1-e^{-2\kappa a_{jk}})$, and
\begin{equation}
{\bf t}_{ijk}^D = 
2\kappa\left[ \begin{array}{ccc}
\dfrac{e^{\kappa a_{ij}}}{2\sinh\kappa a_{ij}} 
& -\dfrac{1}{2\sinh\kappa a_{ij}} & 0 \\[3mm]
-\dfrac{1}{2\sinh\kappa a_{ij}} \hspace{4mm}
& \dfrac{e^{\kappa a_{ij}}}{2\sinh\kappa a_{ij}} 
+ \dfrac{e^{\kappa a_{jk}}}{2\sinh\kappa a_{jk}} -1 \hspace{4mm}
& -\dfrac{1}{2\sinh\kappa a_{jk}} \\[3mm]
0 & -\dfrac{1}{2\sinh\kappa a_{jk}} 
& \dfrac{e^{\kappa a_{jk}}}{2\sinh\kappa a_{jk}}
\end{array}\right]
= {\bf t}_i^D + {\bf t}_j^D + {\bf t}_k^D
+ \Delta {\bf t}_{ij}^D + \Delta {\bf t}_{jk}^D,
\end{equation}
in the limit of three Dirichlet plates.
This implies $\Delta {\bf t}_{ijk}^D +\Delta {\bf t}_{ik}^D = 0$,
and is explicitly verified by Eq.~(\ref{Delt123-3psol})
or Eq.~(\ref{tiDk=-tik}),
\begin{equation}
\Delta {\bf t}_{ijk}^D = -\frac{\kappa}{\sinh\kappa a_{ik}}
\left[ \begin{array}{ccc} e^{-\kappa a_{ik}} & 0 & -1 \\ 
0 & 0 & 0 \\
-1 & 0 & e^{-\kappa a_{ik}}
\end{array}\right] = -\Delta {\bf t}_{ik}^D,
\end{equation}
using Eq.~(\ref{t12D=exp}).

\section{Many-body Casimir energies}
\label{Nb-Cas-En:s}

Casimir energies are finite parts of the vacuum energy that describe
its dependence on configurations of macroscopic objects. The interaction
of classical objects with quantized fields at low energies can be 
described by background potentials. It is 
known~\cite{Deutsch:1978sc, kirsten2002spectral, Fulling:1989nb, Bordag:PRD70.045003, Schaden:2010wv}
that such a semiclassical description for the interaction with a 
quantized field suffers of (local) ultraviolet divergences. A proper
treatment of the interaction at high energies requires modeling of
the quantum fluctuations associated with the objects. One fortunately
sometimes is able to isolate parts of the vacuum energy that depend only on
global changes of the system and can be reliably computed in 
semiclassical approximation. In the following we systematically
determine irreducible parts of the vacuum energy for a
given number of classical objects. These irreducible $N$-body Casimir
energies diverge only if all $N$ potentials describing the classical
objects have a region of common support~\cite{Schaden:2010wv}.

Let $E_0$ be the (infinite) vacuum energy associated with zero-point
fluctuations of a massless scalar field in the absence of all
background potentials, $V_i({\bf x})$. The change in vacuum energy
in the presence of $N$ objects associated with the background potential,
$V=\sum_i V_i$, can be derived using field theory techniques,
for example in \cite{Bordag:1992cm, CaveroPelaez:2008tj}, to be
\begin{equation}
E_{1\ldots N}-E_0 = - \frac{1}{2} \int_{-\infty}^{\infty}\frac{d\zeta}{2\pi}
\,2\zeta^2\, \text{Tr}\, (G_{1\ldots N}-G_0)
= - \frac{1}{2} \int_{-\infty}^{\infty}\frac{d\zeta}{2\pi}
\,\text{Tr}\ln \tilde G_{1\ldots N}
= - \frac{1}{2} \int_{-\infty}^{\infty}\frac{d\zeta}{2\pi}
\,\text{Tr}\ln (1-\tilde T_{1\ldots N}).
\label{TrG-for}
\end{equation}
These expressions have recently been dubbed the Trace-G-formula and 
Trace-Log-G-formula respectively. For frequency independent potentials,
the relation between them is established by differentiating 
Eq.~(\ref{Gi-def}),
\begin{equation}
-\frac{d}{d\zeta^2}\, G = GG,
\label{GG-iden}
\end{equation}
and ignoring a boundary term.

To proceed further we generalize Eqs.~(\ref{G12-casc}) and (\ref{G123-casc})
and decompose a Green's function involving $N$ potentials
into irreducible $N$-body parts,
\begin{equation}
G_{1\ldots N} = G_0 + \sum_i\Delta G_i + \sum_{i<j}
\Delta G_{ij} + \dots.
\label{GN-casc}
\end{equation}
Using Eq.~(\ref{GG-iden}), the irreducible one--two--and three-body
parts of the Green's functions can be written in the form ($i\neq j\neq k$)
\begin{subequations}
\begin{eqnarray}
\Delta G_{i} &=& G_i-G_0 = -\frac{d}{d\zeta^2}\,\ln \frac{G_{i}}{G_0}, \\
\Delta G_{ij} &=& G_{ij}-\Delta G_i - \Delta G_j
=-\frac{d}{d\zeta^2}\,\ln \frac{G_{ij}G_0}{G_iG_j}, \\
\Delta G_{ijk} &=& G_{ijk} - \Delta G_{ij}- \Delta G_{jk}- \Delta G_{ki}
-\Delta G_i - \Delta G_j - \Delta G_k =-\frac{d}{d\zeta^2}\,\ln
\frac{G_{ijk}G_iG_jG_k}{G_{ij}G_{jk}G_{ki}G_0},
\end{eqnarray}
\end{subequations}
which is a (cascading) recursive definition 
that can be extended to higher $N$.

Eqs.~(\ref{TrG-for}) and (\ref{GN-casc}) imply a corresponding decomposition
of the vacuum energy into irreducible $N$-body contributions,
\begin{equation}
E_{1\ldots N} = E_0 + \sum_i\Delta E_i + \sum_{i<j} \Delta E_{ij} + \dots,
\label{EN-casc}
\end{equation}
where
\begin{equation}
\Delta E_{1\ldots N} = - \frac{1}{2} \int_{-\infty}^{\infty}\frac{d\zeta}{2\pi}
\,2\zeta^2\, \text{Tr}\, \Delta G_{1\ldots N}.
\label{DE1N=DG1N}
\end{equation}
As shown in~\cite{Schaden:2010wv}, and as will be explicitly verified in 
examples below, the irreducible $N$-body contribution to the vacuum energy
diverges only if all $N$ potentials have a common support. 
One-body vacuum energies thus are generically divergent, whereas 
two-body Casimir energies diverge only if the two bodies intersect 
(and thus could be viewed as one). More interestingly though, 
three-body Casimir energies diverge only when all three objects 
have a \emph{common} intersection--the three bodies need not be 
mutually disjoint and could, for instance, be arranged to form a triangle.

Eq.~(\ref{G1-N=M1N}) relates the irreducible $N$-body contribution
of the Green's functions to the irreducible $N$-body transition matrix,
\begin{equation}
\text{Tr}\, \Delta G_{1\ldots N} =-\text{Tr}\, \Delta T_{1\ldots N}G_0 G_0
=\text{Tr}\, \Delta T_{1\ldots N} \frac{d}{d\zeta^2} G_0.
\label{GN-TN}
\end{equation}
The support of delta-function potentials $V_i$ is restricted to 
the surface $S_i$ of the $i$-th object and components of the transition-matrix
at most have support on the union of two such surfaces. 
It is therefore convenient to formally define a vector ${\bf R}({\bf x})$,
and a matrix ${\bf R}$, with components
\begin{equation}
{\bf R}_{i}({\bf x}) :=G_0({\bf x}-{\bf y})\big|_{{\bf y}\in S_i},
\qquad {\bf R}_{ij}:= 
G_0({\bf x}-{\bf y})\big|_{\genfrac{}{}{0pt}{}{{\bf x}\in S_i}{{\bf y}\in S_j}}.
\label{def-R}
\end{equation}
Using these definitions in Eq.~(\ref{GG-iden}) we have
\begin{equation}
-\frac{d}{d\zeta^2} {\bf R} 
=\int d^3x\, {\bf R}({\bf x})\cdot {\bf R}({\bf x})^T.
\label{dR=RR}
\end{equation}
Writing the irreducible $N$-body transition
operator in Eq.~(\ref{GN-TN}) in matrix notation and using 
Eq.~(\ref{dR=RR}), we express the irreducible
$N$-body contribution to the vacuum energy of Eq.~(\ref{DE1N=DG1N})
in the form
\begin{equation}
\Delta E_{1\ldots N} = -\frac{1}{2} \int_{-\infty}^{\infty}\frac{d\zeta}{2\pi}
\,2\zeta^2\, \text{Tr}
\left[ \Delta {\bf T}_{1\ldots N} \cdot \frac{d}{d\zeta^2} {\bf R} \right].
\label{Tr-TR-for}
\end{equation}
The trace in the last expression is over matrix indices and includes 
integrals over the lower dimensional surfaces of the associated objects.
Note also that Eq.~(\ref{Tr-TR-for}) 
involves the irreducible $N$-body transition matrix, ${\bf T}_{1\ldots N}$, 
not its partly amputated cousin $\tilde{\bf T}_{1\ldots N}$.

\section{Casimir energies for parallel semitransparent $\delta$-plates}
\label{Cas-en-semiP:s}

We illustrate this formalism by evaluating the irreducible (scalar)
Casimir energy for semitransparent parallel plates described by potentials 
of the form given in Eq.~(\ref{Vi-def}). Exploiting translational 
invariance parallel to the plates in Eq.~(\ref{Tr-TR-for}),
the irreducible $N$-body Casimir energy per unit area is described by
dimensionally reduced quantities 
\begin{equation}
\frac{\Delta E_{1\ldots N}}{L_xL_y} 
=-\frac{1}{6\pi^2} \int_0^\infty \kappa^4d\kappa \,\text{Tr} 
\left[ \Delta {\bf t}_{1\ldots N} \cdot \frac{d}{d\zeta^2} {\bf r} \right],
\label{Tr-TR-forp}
\end{equation}
where $L_x$ and $L_y$ are the (infinite) lengths of the plates in 
$x$ and $y$ direction, respectively. 
The integrals on $\zeta$, $k_x$, and $k_y$, are performed using 
polar varibales, which effectively amounts to replacing
$\zeta^2\to 2\kappa^2/3$, where $\kappa$ was defined after Eq.~(\ref{Gi=Igi}).
The dimensionally reduced transition matrices,
$\Delta {\bf t}_{1\ldots N}$, for $N=2$ and $N=3$ are, respectively,
given by Eqs.~(\ref{Dt12=sol}) and (\ref{Delt123-3psol}). 
The derivative of the dimensionally reduced free Green's function in 
this case is the matrix
\begin{equation}
-\frac{d}{d\zeta^2} {\bf r} =\int_{-\infty}^\infty dz\, 
{\bf r}(z)\cdot {\bf r}(z)^T
= \frac{2}{(2\kappa)^3} \left[ \begin{array}{cccc}
1 & (1+\kappa a_{12})\,e^{-\kappa a_{12}} & \cdots
& (1+\kappa a_{1N})\,e^{-\kappa a_{1N}} \\[1mm]
(1+\kappa a_{21})\,e^{-\kappa a_{21}} &1 & \cdots
& (1+\kappa a_{2N})\,e^{-\kappa a_{2N}} \\[1mm]
\vdots & \vdots & \ddots & \vdots \\[1mm]
(1+\kappa a_{N1})\,e^{-\kappa a_{N1}} & (1+\kappa a_{N2})\,e^{-\kappa a_{N2}} 
& \cdots & 1 
\end{array} \right],
\label{pp-nmat}
\end{equation}
where $a_{ij}$ is the distance between the $i$-th and $j$-th parallel plate
defined previously.

\subsection{$N=1,2$: Irreducible one--and two-body Casimir energy for 
semitransparent plates}

The irreducible one-body vacuum energy per unit area associated with 
the $i$-th plate diverges. Eq.~(\ref{Tr-TR-forp}) gives it as the integral
\begin{equation}
\frac{\Delta E_i}{L_xL_y} 
=\frac{1}{12\pi^2}\int_0^\infty\kappa^2 d\kappa\,\bar{t}_i,
\end{equation}
with $\bar{t}_i$ defined in Eq.~(\ref{g1-exp}).
The one-body vacuum energies are ultra-violet divergent at any 
non-vanishing coupling, but do not depend on the relative position 
of the plates and therefore do not contribute to forces between them.

The irreducible two-body Casimir energy per unit area associated with plates 
$i$ and $j$ is obtained by inserting 
Eqs.~(\ref{Dt12=sol}) and (\ref{pp-nmat}) (for $N=2$) 
in Eq.~(\ref{Tr-TR-forp}),
\begin{equation}
\frac{\Delta E_{ij}}{L_xL_y} 
= - \frac{1}{12\pi^2} \int_0^\infty \kappa^2d\kappa
\left[ \frac{1}{\Delta_{ij}} -1 \right]
\big[ 2\kappa a_{ij} + (1-\bar{t}_i) + (1-\bar{t}_j) \big],
\label{2-semi}
\end{equation}
where the two-body determinant is given by
Eq.~(\ref{Xij-sol}). Eq.~(\ref{2-semi}) for the Casimir interaction 
energy of two semitransparent plates was obtained previously 
in \cite{Bordag:1992cm}.
In the Dirichlet limit, $\bar{t}_i\to 1$, Eq.~(\ref{2-semi}) simplifies
to the well known Casimir energy for a massless scalar field satisfying
Dirichlet boundary conditions on a pair of parallel plates,
\begin{equation}
\frac{\Delta E_{ij}^D}{L_xL_y} 
= - \frac{1}{12\pi^2} \int_0^\infty \kappa^2d\kappa
\frac{2\kappa a_{ij}}{e^{2\kappa a_{ij}}-1}
=-\frac{\pi^2}{1440}\frac{1}{a_{ij}^3}.
\label{DCasplates}
\end{equation}
Eqs.~(\ref{2-semi}) and (\ref{DCasplates}) are finite and negative for 
two disjoint plates.

\subsection{$N=3$: Three-body Casimir energy for three parallel plates}

The irreducible three-body Casimir energy of three plates is
similarly obtained by inserting Eqs.~(\ref{Delt123-3psol}) and (\ref{pp-nmat})
(for $N=3$) in Eq.~(\ref{Tr-TR-forp}),
\begin{eqnarray}
\frac{\Delta E_{ijk}}{L_xL_y} 
&=& \frac{1}{12\pi^2} \int_0^\infty \kappa^2d\kappa
\bigg[ 
\bar{t}_i \left\{ 1 -\frac{1}{\Delta_{ij}} -\frac{1}{\Delta_{ik}} 
+\frac{\Delta_{jk}}{\Delta_{ijk}} \right\} 
+ 2(1+\kappa a_{jk}) \left\{\frac{1}{\Delta_{jk}} -1\right\} 
\left\{1-\bar r_{jk[i]}e^{\kappa a_{jk}}\frac{\Delta_{jk}}{\Delta_{ijk}}\right\}
\nonumber \\ && \hspace{25mm}
+ \bar{t}_j \left\{ 1 -\frac{1}{\Delta_{ji}} -\frac{1}{\Delta_{jk}} 
+\frac{\Delta_{ik}}{\Delta_{ijk}} \right\} 
+ 2(1+\kappa a_{ik}) \left\{\frac{1}{\Delta_{ik}} -1\right\} 
\left\{1-\bar r_{ik[j]}e^{\kappa a_{ik}}\frac{\Delta_{ik}}{\Delta_{ijk}}\right\}
\nonumber \\ && \hspace{25mm}
+\bar{t}_k \left\{ 1 -\frac{1}{\Delta_{ki}} -\frac{1}{\Delta_{kj}} 
+\frac{\Delta_{ij}}{\Delta_{ijk}} \right\}
+ 2(1+\kappa a_{ij}) \left\{\frac{1}{\Delta_{ij}} -1\right\} 
\left\{1-\bar r_{ij[k]}e^{\kappa a_{ij}}\frac{\Delta_{ij}}{\Delta_{ijk}}\right\}
\bigg].
\label{DE123=gen}
\end{eqnarray}

When Dirichlet boundary conditions are imposed on the central $j$-th plate,
the relation between irreducible two--and three-body transition matrices 
noted in Eq.~(\ref{etijk=2tik}) implies a corresponding relation 
between two--and three-body Casimir energies,
\begin{equation}
\Delta E_{i\infty k} + \Delta E_{ik} = 0.
\end{equation}
This is explicitly verified by using the factorization of the 
three-body determinant in Eq.~(\ref{det-fact}) and the Dirichlet limits for 
$\bar r_{ij[k]}$ given after Eq.~(\ref{det-fact}) in Eq.~(\ref{DE123=gen}), 
and identifying the irreducible two-body energy of Eq.~(\ref{2-semi})
in the result.

In the Dirichlet limit for all three plates the irreducible three-body
Casimir energy cancels the well-known two-body interaction between the
outer Dirichlet plates,
\begin{equation}
\frac{\Delta E_{ijk}^D}{L_xL_y} =\frac{\pi^2}{1440}\frac{1}{a_{ik}^3}
= -\frac{\Delta E_{ik}^D}{L_xL_y},
\end{equation}
where $a_{ik}$ is the distance between the outer plates.

This cancellation is to be expected on physical grounds and serves 
to check the calculation. For semitransparent plates the cancellation
is not complete and the irreducible three-body contribution to
the total Casimir energy can be significant
for parallel plates. Note that the sign of the irreducible $N$-body 
contribution to the scalar Casimir energy alternates.
Although not apparent from the expression of Eq.~(\ref{DE123=gen}),
this irreducible three-body
contribution to the Casimir energy is positive for \emph{any} positive
couplings $\lambda_1,\lambda_2,\lambda_3$ and \emph{any} relative 
position of the three plates. 
For parallel semitransparent plates we thus verify the more 
general result obtained in~\cite{Schaden:2010wv}.
Also, as discussed in~\cite{Schaden:2010wv}
and noted previously, the three-plate Casimir energy diverges only if
\emph{all three} plates coincide.

In the following we will see that these generic results for the sign
and analyticity of the three-body scalar Casimir energy hold in the
limit where two of the three potentials are weak and need only be
accounted for to leading order.

\section{Three-body Scalar Casimir interaction for semiweak coupling}
\label{3body-sweak:s}

We now consider irreducible vacuum energies for three  bodies when 
two of the three potentials, $V_1$ and $V_2$, 
are weak and need only be taken to leading
order. No restriction is imposed on the potential $V_3$ describing the
third body. To the leading order we thus approximate 
$T_1\sim V_1$ and $T_2\sim V_2$ in Eq.~(\ref{ti-def}).
The three-body transition matrix of Eq.~(\ref{123-rtr}) 
in this semiweak approximation simplifies to
\begin{equation}
\tilde {\bf T}_{123}^W =
\left[ \begin{array}{ccc}
1 & 0&0 \\ 0&1&0 \\ 0&0& X_{3[12]}^W
\end{array}\right]
\left[ \begin{array}{ccc}
\tilde V_1 & -\tilde V_1(1-\tilde T_3)\tilde V_2 
& -\tilde V_1(1-\tilde V_2+\tilde T_3\tilde V_2)\tilde T_3 \\
-\tilde V_2(1-\tilde T_3)\tilde V_1 & \tilde V_2 
& -\tilde V_2(1-\tilde V_1+\tilde T_3\tilde V_1)\tilde T_3 \\
-\tilde T_3(1-\tilde V_2)\tilde V_1 
& -\tilde T_3(1-\tilde V_1)\tilde V_2 & \tilde T_3
\end{array}\right].
\label{123-rtr-w}
\end{equation}
Here $X_{3[12]}^W$ satisfies Eq.~(\ref{Xijk-def}), 
which to leading semiweak approximation is solved by
\begin{equation}
X_{3[12]}^W = 1 +\tilde T_3(1-\tilde V_1)\tilde V_2 
+\tilde T_3(1-\tilde V_2)\tilde V_1 
+\tilde T_3\tilde V_1\tilde T_3\tilde V_2 
+\tilde T_3\tilde V_2\tilde T_3\tilde V_1.
\end{equation}
The transition matrix in semiweak approximation of Eq.~(\ref{123-rtr-w})
may again be decomposed into its irreducible 
one--two--and three-body parts, leading to the semiweak version of 
Eq.~(\ref{T123=T1..DT123}).
From Eq.~(\ref{DT12=expl}) the irreducible two-body transition matrices in
semiweak approximation are,
\begin{equation}
\Delta \tilde{\bf T}_{12}^W= 
\left[ \begin{array}{cc} 0 & -\tilde V_1\tilde V_2 \\ 
-\tilde V_2\tilde V_1&0 \end{array}\right],
\qquad \Delta \tilde {\bf T}_{i3}^W= 
\left[ \begin{array}{cc} 0 & -\tilde V_i\tilde T_3 \\ 
-\tilde T_3\tilde V_i& \tilde T_3\tilde V_i\tilde T_3 \end{array}\right],
\label{TW12}
\end{equation}
with $i=1,2$. Similarly approximating Eq.~(\ref{d123-rtr}), 
the three-body transition matrix in semiweak approximation becomes,
\begin{equation}
\Delta \tilde{\bf T}_{123}^W
= \left[ \begin{array}{ccc}
0 & \tilde V_1\tilde T_3\tilde V_2 
& \tilde V_1\tilde G_3\tilde V_2\tilde T_3 \\ 
\tilde V_2\tilde T_3\tilde V_1&0 & \tilde V_2\tilde G_3\tilde V_1\tilde T_3 \\
\tilde T_3\tilde V_2\tilde G_3\tilde V_1 \hspace{3mm} 
& \tilde T_3\tilde V_1\tilde G_3\tilde V_2 \hspace{3mm} & 
-\tilde T_3\tilde V_1\tilde G_3\tilde V_2\tilde T_3 
-\tilde T_3\tilde V_2\tilde G_3\tilde V_1\tilde T_3
\end{array}\right],
\label{DT123W}
\end{equation}
where $\tilde G_3=1-\tilde T_3$.

Casimir energies in the semiweak approximation are obtained using 
Eqs.~(\ref{DE1N=DG1N}) and (\ref{GN-TN}). Inserting Eq.~(\ref{TW12}) 
in Eq.~(\ref{GN-TN}) we have to this approximation,
\begin{subequations}
\begin{eqnarray}
-\text{Tr}\,\Delta G_{12}^W = \text{Tr} 
\left[ \Delta {\bf T}_{12}^W \cdot \frac{d}{d\zeta^2} {\bf R} \right]
&=& \frac{d}{d\zeta^2} \text{Tr} \Big[ G_0V_1G_0V_2 \Big], \\
-\text{Tr}\,\Delta G_{i3}^W = \text{Tr}
\left[ \Delta {\bf T}_{i3}^W \cdot \frac{d}{d\zeta^2} {\bf R} \right]
&=& \frac{d}{d\zeta^2} \text{Tr} \Big[ G_0V_iG_0T_3 \Big], \qquad (i=1,2).
\end{eqnarray}
\label{DG12=dztr}
\end{subequations}
The corresponding irreducible three-body contribution 
using Eq.~(\ref{DT123W}) in Eq.~(\ref{GN-TN}) is
\begin{equation}
-\text{Tr}\,\Delta G_{123}^W = \text{Tr}
\left[ \Delta {\bf T}_{123}^W \cdot \frac{d}{d\zeta^2} {\bf R} \right]
= -\frac{d}{d\zeta^2} \text{Tr} \Big[ 
G_0V_1G_0T_3G_0V_2 +G_0V_2G_0T_3G_0V_1 - G_0T_3G_0V_1G_0T_3G_0V_2 \Big].
\label{TrDT123R}
\end{equation}
Inserting Eq.~(\ref{DG12=dztr}) in Eq.~(\ref{DE1N=DG1N}),
and integrating by parts, the irreducible two-body Casimir energies
in semiweak approximation are 
\begin{subequations}
\begin{eqnarray}
\Delta E_{12}^W &=& -\frac{1}{2} \int_{-\infty}^{\infty}\frac{d\zeta}{2\pi}
\,\text{Tr} \big[G_0V_1G_0V_2\big], \\ 
\Delta E_{i3}^W &=& -\frac{1}{2} \int_{-\infty}^{\infty}\frac{d\zeta}{2\pi}
\,\text{Tr} \big[G_0V_iG_0T_3\big], \qquad (i=1,2), 
\label{DelE=2VVT-w}
\end{eqnarray}
\label{DelE=2VVT-wcom}%
\end{subequations}
verifying results reported in \cite{Milton:2007wz}.
The corresponding irreducible three-body contribution to the Casimir energy
in semiweak approximation
using Eq.~(\ref{TrDT123R}) in Eq.~(\ref{DE1N=DG1N}) is
\begin{equation}
\Delta E_{123}^W = \frac{1}{2} \int_{-\infty}^{\infty}\frac{d\zeta}{2\pi}
\,\text{Tr} \big[
G_0V_1G_0T_3G_0V_2 +G_0V_2G_0T_3G_0V_1 - G_0T_3G_0V_1G_0T_3G_0V_2 \big].
\label{DelE=VVT-w}
\end{equation}

In the following we evaluate Eqs.~(\ref{DelE=2VVT-wcom}) and
(\ref{DelE=VVT-w}) for some special cases.


\subsection{Weak point potentials}

Weak point potentials of the form,
\begin{equation}
V_i({\bf x})=\lambda_i\delta({\bf x}-{\bf x}_i),
\end{equation}
for $i=1,2$, allow one to explicitly perform the integrals
in Eqs.~(\ref{DelE=2VVT-wcom}) and (\ref{DelE=VVT-w}).
In this case we have
\begin{equation}
\Delta E_{12}^W 
= -\frac{\lambda_1\lambda_2}{2} \int_{-\infty}^{\infty}\frac{d\zeta}{2\pi}
\,\big[ G_0({\bf x}_1-{\bf x}_2)\big]^2<0, 
\end{equation}
and, using Eq.~(\ref{Gi-gsol}) in Eq.~(\ref{DelE=2VVT-w}),
\begin{equation}
\Delta E_{i3}^W 
= -\frac{\lambda_i}{2} \int_{-\infty}^{\infty}\frac{d\zeta}{2\pi}
\,\Big\{ G_0(0)-G_3({\bf x}_i,{\bf x}_i)\Big\}<0, \qquad (i=1,2),
\label{2point-w}
\end{equation}
because the integrand in braces is positive for positive $V_3$.
The irreducible two-body contributions to the vacuum energy thus are negative
for weak point potentials. We similarly obtain that the irreducible three-body
correction to the vacuum energy,
\begin{equation}
\Delta E_{123}^W
= \frac{\lambda_1\lambda_2}{2} \int_{-\infty}^{\infty}\frac{d\zeta}{2\pi}
\,\Big\{ \big[G_0({\bf x}_1-{\bf x}_2)\big]^2
-\big[G_3({\bf x}_1,{\bf x}_2)\big]^2 \Big\}>0,
\label{3point-w}
\end{equation}
in this case is positive for any (positive) potential $V_3$.
Note that the irreducible three-body Casimir energy in semiweak 
approximation diverges only if ${\bf x}_1={\bf x}_2$ is in the 
support of $V_3$.

The pattern in the sign of the irreducible $N$-body contribution 
is consistent with the findings of \cite{Schaden:2010wv}. Furthermore,
since any positive potential is a (positive) superposition of point
potentials, this pattern of the signs of irreducible $N$-body
contributions extend to any shape of the objects in semiweak approximation.
This is explicitly verified by the following examples.

\subsection{Weak potentials with translational symmetry 
parallel to a Dirichlet plate}

\begin{center}
\begin{figure}
\includegraphics{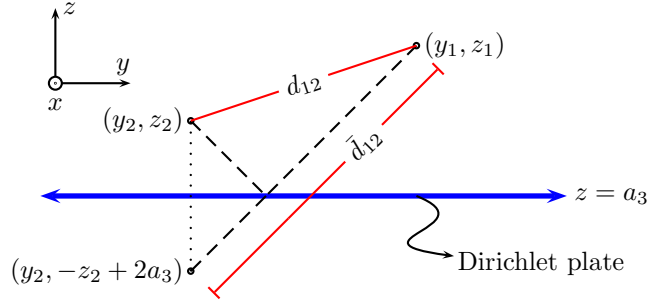}
\caption{The distances $d_{12}$ and $\bar{d}_{12}$.
The effective distance
$\bar{d}_{12}$ is the shortest distance between the two points for a path 
that reflects off the Dirichlet plate at $z=a_3$. 
It also is the shortest distance between $(y_1,z_1)$ and a mirror
image of the point $(y_2,z_2)$ with respect to the $z=a_3$ line.}
\label{r12-rb12}
\end{figure}
\end{center}

We consider a Dirichlet plate and weak potentials that do not depend 
on the Cartesian coordinate $x$,
\begin{equation}
V_i=V_i(y,z), \quad \text{for} \quad i=1,2;
\quad \text{and} \quad
V_3=\lambda_3\,\delta(z-a_3), \quad \text{with} \quad \lambda_3\to\infty.
\end{equation}
To evaluate Eqs.~(\ref{DelE=2VVT-wcom}) and (\ref{DelE=VVT-w})
for such potentials
we require the operator $G_0T_3^DG_0$ for a Dirichlet plate.
In order to exploit the translational symmetry in $x$-direction
we write the solution to Eq.~(\ref{G0-def}) for the free Green's function
in the form
\begin{equation}
G_0(|{\bf x}_1-{\bf x}_2|;\zeta)
= \int \frac{d^2k}{(2\pi)^2}\,e^{i{\bf k}\cdot({\bf x}_1-{\bf x}_2)_\perp}
\frac{e^{-\kappa |z_1-z_2|}}{2\kappa}
= \int_{-\infty}^{\infty}\frac{dk_x}{2\pi}\,e^{ik_x(x_1-x_2)}
\frac{K_0(\bar{\kappa}\, d_{12})}{2\pi}
=\frac{e^{-|\zeta| |{\bf x}_1-{\bf x}_2|}}{4\pi |{\bf x}_1-{\bf x}_2|},
\label{G0=K0}
\end{equation}
where $d_{12}=\sqrt{(y_1-y_2)^2+(z_1-z_2)^2}$ is the projected distance
in the $x_1=x_2$ plane, and $\bar{\kappa}^2= k_x^2+\zeta^2$.
$K_0(x)$ is the modified Bessel function of zero order.
Note that $\kappa$ defined after Eq.~(\ref{Gi=Igi}) satisfies 
$\kappa^2 = \bar{\kappa}^2+k_y^2$.
Using the first equality of Eq.~(\ref{G0=K0})
and the dimensionally reduced transition matrix of a Dirichlet plate 
given in Eq.~(\ref{g1-exp}) one can show that 
\begin{equation}
-\Delta G_3^D({\bf x}_1,{\bf x}_2;\zeta)
=[G_0T_3^DG_0]({\bf x}_1,{\bf x}_2;\zeta)
= G_0(|{\bf x}_1-\bar{\bf x}_2|;\zeta)
= \int_{-\infty}^{\infty}\frac{dk_x}{2\pi}\,e^{ik_x(x_1-x_2)}
\frac{K_0(\bar{\kappa}\, \bar{d}_{12})}{2\pi},
\label{G0T3DG0}
\end{equation}
where $\bar{\bf x}_2=(x_2,y_2,-z_2+2a_3)$, and
$\bar{d}_{12}$ is the length of the shortest path between 
${\bf x}_1$ and ${\bf x}_2$ in the $(x_1=x_2)$-plane 
that reflects off the Dirichlet plate. 
For a Dirichlet plate at $z=a_3$, this distance is given by 
$\bar{d}^2_{12}=(y_1-y_2)^2+(|z_1-a_3|+|z_2-a_3|)^2$. 
A geometrical interpretation of $d_{12}$ and $\bar{d}_{12}$ 
is shown in FIG.~\ref{r12-rb12}.
Substituting Eq.~(\ref{G0T3DG0}) in Eq.~(\ref{Gi-gsol}) leads to
\begin{equation}
G_3^D({\bf x}_1,{\bf x}_2;\zeta)
=G_0(|{\bf x}_1-{\bf x}_2|;\zeta)- G_0(|{\bf x}_1-\bar{\bf x}_2|;\zeta),
\label{G1D=G0G0}
\end{equation}
which is anti-symmetric under reflection about the Dirichlet plate. 
Note that if ${\bf x}_1$ and ${\bf x}_2$ are on opposite
sides of the plate, $\bar{d}_{12}=d_{12}$, and $G_3^D$ vanishes.

Substituting Eq.~(\ref{G0=K0}) for the free Green's functions,
and Eq.~(\ref{G0T3DG0}) for the irreducible Green's function of a Dirichlet
plate in Eqs.~(\ref{DelE=2VVT-wcom}) and using the identity
\begin{equation}
\int_0^\infty \bar{\kappa}\,d\bar{\kappa} \,K_0(\bar{\kappa} x) 
= \frac{1}{x^2},
\label{K0-iden}
\end{equation}
the irreducible two-body Casimir energies per unit length
in semiweak approximation for potentials with translational symmetry are
\begin{subequations}
\begin{eqnarray}
\frac{\Delta E_{12}^W}{L_x} 
&=& -\frac{1}{32\pi^3} \int d^2r_1 \int d^2r_2 
\frac{V_1({\bf r}_1) V_2({\bf r}_2)}{d_{12}^2}, 
\label{V1V2-w} \\[2mm] 
\frac{\Delta E_{i3}^W}{L_x} 
&=& -\frac{1}{32\pi^2} \int d^2r \frac{V_i({\bf r})}{|z|^2}, 
\qquad (i=1,2).
\label{E12W-ViDP}
\end{eqnarray}
\label{E12W-all}%
\end{subequations}
The Casimir energy in Eq.~(\ref{V1V2-w}) for two weakly interacting 
objects with translational symmetry was previously obtained in
\cite{Wagner:2008qq}.
The Casimir energy for a Dirichlet plate weakly interacting with an
object possessing translational symmetry was obtained in \cite{Milton:2007wz},
but was given as a series involving modified Bessel functions.
The expression in \cite{Milton:2007wz} generally is much harder to 
evaluate than Eq.~(\ref{E12W-ViDP}).
The simplification in Eq.~(\ref{E12W-ViDP}) was achieved by 
using the effective Green's function for a Dirichlet plate in 
Eq.~(\ref{G0T3DG0}). For many potentials, the evaluation of the
Casimir energy by Eq.~(\ref{E12W-ViDP}) is immediate.
We can for example 
calculate the two-body Casimir energy for a cylinder of radius $a$,
described by the weak potential $V_\text{cyl}=\lambda\,\delta (r-a)$,
interacting with a Dirichlet plate positioned at $z=R>a$. 
From Eq.~(\ref{E12W-ViDP}) one readily finds,
\begin{equation}
\frac{\Delta E_\text{Cyl-DP}^W}{L_x}
= -\frac{1}{32\pi^2} \int_0^\infty rdr \int_0^{2\pi} d\theta
\frac{\lambda\delta (r-a)}{|r\sin\theta -R|^2}
=-\frac{\lambda a}{16\pi}\frac{1}{R^2}
\left[1 - \frac{a^2}{R^2} \right]^{-\frac{3}{2}},
\end{equation}
which reproduces the expression in \cite{Milton:2007wz}.
A similarly simplified evaluation is expected for
an arbitrary surface with translational symmetry weakly interacting
with a Dirichlet plate parallel to the symmetry axis.

The irreducible three-body Casimir energies for translationally 
invariant weak potentials
and a Dirichlet plate can be similarly evaluated using
Eq.~(\ref{DelE=VVT-w}).
The first two terms in Eq.~(\ref{DelE=VVT-w}) involve the product of 
the free Green's function, $G_0$, with the irreducible Green's function 
for a Dirichlet plate given in Eq.~(\ref{G0T3DG0}).
The last term requires the product of two irreducible one-body Green's
functions. A useful identity for the product of two modified Bessel functions
of zeroth order is
\begin{equation}
\int_0^\infty \bar{\kappa}\,d\bar{\kappa}
\,K_0(\bar{\kappa} x) K_0(\bar{\kappa} y)
= \frac{1}{x^2-y^2} \ln \left(\frac{x}{y}\right)
\xrightarrow{x\to y} \frac{1}{2x^2}.
\label{K0-iden2}
\end{equation}
Inserting Eqs.~(\ref{G0=K0}) and (\ref{G0T3DG0}) in Eq.~(\ref{DelE=VVT-w}) 
to write the Green's functions in terms of modified Bessel functions,
and then using Eq.~(\ref{K0-iden2}), we obtain 
\begin{equation}
\frac{\Delta E_{123}^W}{L_x}
= \frac{1}{32\pi^3} \int d^2r_1 \int d^2r_2 
\frac{V_1({\bf r}_1) V_2({\bf r}_2)}{\bar{d}_{12}^2}
\;Q\left(\frac{d_{12}^2}{\bar{d}_{12}^2} \right),
\label{DelE=pot-w}
\end{equation}
where the distances $d_{12}$ and $\bar{d}_{12}$ were introduced earlier
and are shown in FIG.~\ref{r12-rb12}. 
The function
\begin{equation}
Q(x) = -\frac{2\ln x}{1-x}-1
\label{Q3-kernel}
\end{equation}
is bounded by $1\leq Q(x)\leq 1-2\ln x$
in the relevant domain $0<x=\frac{d_{12}^2}{\bar{d}_{12}^2}<1$.
This implies that the three-body Casimir energy of Eq.~(\ref{DelE=pot-w})
is always positive and bounded by
\begin{equation}
\frac{1}{32\pi^3} \int d^2r_1 \int d^2r_2 
\frac{V_1({\bf r}_1) V_2({\bf r}_2)}{\bar{d}_{12}^2}
\leq \frac{\Delta E_{123}^W}{L_x}\leq
\frac{1}{32\pi^3} \int d^2r_1 \int d^2r_2     
\frac{V_1({\bf r}_1) V_2({\bf r}_2)}{\bar{d}_{12}^2}
\left[ 1-2\ln\left(\frac{d^2_{12}}{\bar{d}^2_{12}}\right)\right].
\label{DE123W-bounds}
\end{equation}
$\bar{d}_{12}$ is the distance between a point on the first object
and another point on the reflected image of the second object 
(see FIG.~\ref{r12-rb12}). It vanishes only at points where the 
two weak objects \emph{and} the Dirichlet plate are concurrent. 
The irreducible three-body Casimir energy in the semiweak approximation
of Eq.~(\ref{DelE=pot-w}) thus is finite if the three objects have 
no point in common. This contribution in particular does not diverge 
as the objects approach the plate or each other, corroborating the 
findings in \cite{Schaden:2010wv}. Note that the lower bound in 
Eq.~(\ref{DE123W-bounds}) is the two-body Casimir energy between 
weak potentials of Eq.~(\ref{V1V2-w}), but with the reflected object
($d_{12}\to\bar{d}_{12}$) and of opposite sign. The irreducible 
three-body Casimir energy approaches the lower bound for 
$\frac{d_{12}^2}{\bar{d}_{12}^2}\sim 1$ and thus partially cancels 
the irreducible two-body energy if one or both objects approach the 
Dirichlet plate. In fact, if the two weakly interacting objects are 
entirely on opposite sides of the Dirichlet plate, the lower bound 
is achieved because $\bar{d}_{12}=d_{12}$ and the three-body 
Casimir energy cancels the two-body interaction energy between them
precisely.

The following examples demonstrate the finiteness, sign, and analyticity,
of three-body contributions to Casimir energies for cases in which irreducible
one--\emph{and} two-body contributions to the vacuum energy diverge.


\section{Triangular-wedge on a Dirichlet plate}
\label{tri-wedge:s}

\begin{center}
\begin{figure}
\includegraphics{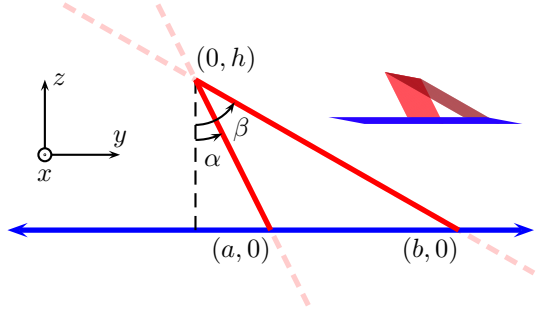}
\caption{Weakly interacting triangular-wedge on a Dirichlet plate.
The objects are of infinite extent in the $x$-direction.
The weakly interacting sides of the wedge (in red) have finite length.}
\label{wedoTri}
\end{figure}
\end{center}

We first consider a triangular-wedge with two sides 
described by weak potentials atop a Dirichlet plate at $z=0$, 
forming a waveguide of triangular cross-section:
\begin{subequations}
\begin{eqnarray}
V_1(y,z) &=& \lambda_1 \delta (-z+m_\alpha(y-a)) \,\theta_1, 
\quad \text{with} \quad 
\theta_1\equiv\theta(y-\text{min}[0,a])\,\theta(\text{max}[0,a]-y), \\
V_2(y,z) &=& \lambda_2 \delta (-z+m_\beta(y-b)) \,\theta_2, 
\quad \text{with} \quad
\theta_2\equiv\theta(y-\text{min}[0,b])\,\theta(\text{max}[0,b]-y),\\
V_3(z) &=& \lambda_3 \delta (z), 
\quad \text{with} \quad \lambda_3\to\infty.
\end{eqnarray}
\label{wed-pot-w}%
\end{subequations}
The sides of the wedge have slopes 
$m_\alpha=-\cot\alpha$ and $m_\beta=-\cot\beta$, 
and lengths $\sqrt{h^2+a^2}$ and $\sqrt{h^2+b^2}$, respectively.
The constraint $m_\alpha a=m_\beta b=-h$ forces the sides to intersect 
at $(y=0,z=h)$, where $h$ is the height of the triangle. 
The base of the triangle formed then measures $|b-a|$.
Note that the Dirichlet plate at $z=0$ is of infinite extent.
This triangular-wedge on a Dirichlet plate is depicted in FIG.~\ref{wedoTri}.
Suitable parameters for describing the triangular waveguide are 
$(h,\alpha,\beta)$, or $(h,\tilde{a}=a/h,\tilde{b}=b/h)$.
Without loss of generality we measure all lengths in multiples of
the height $h$. The triangle then has height $h=1$ and
the parameter space for the triangle is $-\infty<a,b<\infty$,
or, equivalently, $-\pi/2<\alpha,\beta<\pi/2$.

\begin{center}
\begin{figure}
\includegraphics[width=8cm]{figures/3Dplot-E123-versus-alpha-and-beta-for-triangular-wedge-on-Dplate.eps-from-jpg}%
\hspace{10mm}
\includegraphics{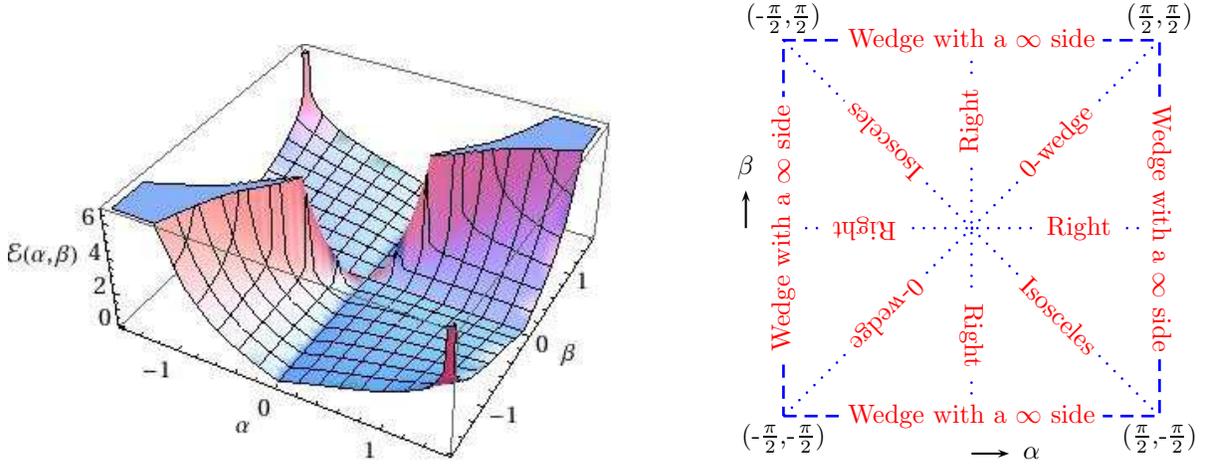}
\caption{ Casimir Landscape: 
${\cal E}(\alpha,\beta)$ as a function of the opening angles
$\alpha$ and $\beta$ for a weakly interacting triangular-wedge 
on a Dirichlet plate. The valley connecting the $\alpha>\beta$ region
with the $\alpha<\beta$ region is an artifact caused by limited numerical
accuracy. The valley should be replaced by a very thin
and infinitely high wall describing
the sharp change in energy when all surfaces overlap. On the right,
the shapes of the triangles are matched to the respective regions 
of the $\alpha$-$\beta$ plane.}
\label{3dTri}
\end{figure}
\end{center}

Observe that all irreducible two-body Casimir energies in Eq.~(\ref{E12W-all})
diverge due to ultra-violet contributions from the corners of the triangle
where pairs of potentials overlap.
More precisely, the integrand $\Delta E_{12}^W$ diverges 
when $d_{12}\sim 0$ near the vertex of the wedge. The integrand of
$\Delta E_{i3}^W$ diverges when $z_i\sim 0$ near the corner with 
the Dirichlet plate. The irreducible three-body Casimir energy,
$\Delta E_{123}^W$, in Eq.~(\ref{DelE=pot-w})
on the other hand is finite because $\bar{d}_{12}$
never vanishes in the integration region.
Substituting the potentials of Eq.~(\ref{wed-pot-w})
for the semiweak triangular waveguide in Eq.~(\ref{DelE=pot-w}) and
evaluating the $z$-integrals gives,
\begin{equation}
{\cal E}(\alpha,\beta)=
\frac{\Delta E_{123}^W}{L_x}
\left[ \frac{\lambda_1\lambda_2}{32\pi^3}\right]^{-1}
=|\tilde{a}\,\tilde{b}|
\int_0^1 \int_0^1 \frac{du_1 du_2}{\bar{u}_{12}^2}
\,Q\left( \frac{u_{12}^2}{\bar{u}_{12}^2} \right),
\label{E123=tri-wed}
\end{equation}
where we have rescaled the integration variables, 
$y_1=|a|u_1$ and $y_2=|b|u_2$ by the respective lengths.
All distances have been expressed in units of $h$:
$d_{12}=hu_{12}$ and $\bar{d}_{12}=h\bar{u}_{12}$, with
\begin{subequations}
\begin{eqnarray}
\bar{u}_{12}^2 &=& (\tilde{a} u_1-\tilde{b} u_2 )^2 + [|1-u_1| +|1-u_2|]^2,\\
u_{12}^2 &=& (\tilde{a} u_1-\tilde{b} u_2)^2 + (u_1 -u_2)^2.
\end{eqnarray}
\end{subequations}
With the function $Q(x)$ defined in Eq.~(\ref{Q3-kernel})
the three-body interaction energy of Eq.~(\ref{E123=tri-wed})
is finite and can be evaluated numerically. 
In FIG.~\ref{3dTri} we plot ${\cal E}(\alpha,\beta)$
as a function of the angles $\alpha$ and $\beta$. 
The three-body interaction energy is always positive and vanishes
(and is minimized) only for $\alpha=0$, or $\beta=0$.
It is minimal when the shorter side of the wedge is perpendicular 
to the Dirichlet plate. Wedges with angles
$\beta<0<\alpha$ or $\alpha<0<\beta$ are energetically 
preferred over wedges with angles $\alpha,\beta>0$ or $\alpha,\beta<0$.
The three-body Casimir-energy diverges only when all three sides of 
the triangle have a point in common, i.e.
when $\alpha =\beta$, or $\alpha=-\beta=\pm\pi/2$.

\begin{center}
\begin{figure}
\includegraphics[width=7.0cm]{figures/3Dplot-E123-versus-area-and-perimeter-for-triangular-wedge-on-Dplate.eps-from-jpeg}
\hspace{10mm}
\includegraphics[width=7.5cm]{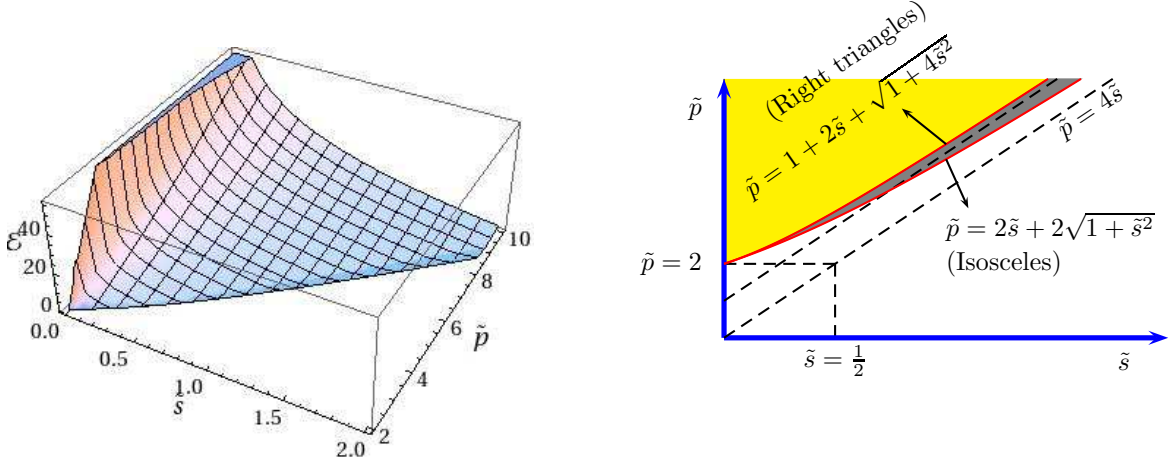}%
\caption{Irreducible three-body Casimir energy of a semiweak
triangular waveguide as a function of the cross-sectional area
and perimeter. This energy vanishes and is minimal along the line
$\tilde{p}=1+2\tilde{s}+\sqrt{1+4\tilde{s}^2}$.}
\label{3dspTri-wed}
\end{figure}
\end{center}

Abalo, Milton, and Kaplan, recently~\cite{Abalo:2010ah} investigated 
the dependence of the Casimir energy on the area and perimeter of 
triangular waveguides on which Dirichlet boundary conditions were imposed.
Although only interior modes were taken into account and divergences 
associated with corners and single-body vacuum energies were removed ad hoc,
they found that the dimensionless Casimir energy of their triangular 
wave guides closely follow a universal function of the dimensionless
ratio ($P^2/A$) of the perimeter $P$ and area $A$ of the cross-section.
This would imply that the Casimir energy of triangular wave guides
depends on just one, rather than two, dimensionless parameters.
Although we cannot expect a similar dependence, the universal behavior
observed in \cite{Abalo:2010ah} prompted us to also investigate the 
dependence of the semiweak three-body Casimir energy on the 
dimensionless perimeter $\tilde p=(P/h)$ and dimensionless area 
$\tilde s=(A/h^2)$ of the triangular waveguide.
It is also of interest to inquire for what configuration the energy of 
a triangular waveguide is minimized if the cross-sectional area
is kept fixed. 
The dimensionless area $\tilde s$ and perimeter $\tilde p$ of the 
triangular wedge are given by,
\begin{subequations}
\begin{eqnarray}
\frac{A}{h^2}&=&\tilde{s}=\frac{1}{2}|\tilde{b}-\tilde{a}|,\label{s=ab} \\
\frac{P}{h} &=& \tilde{p}= |\tilde{b}-\tilde{a}| 
+ \sqrt{1+\tilde{a}^2} +\sqrt{1+\tilde{b}^2}.%
\label{p=ab}
\end{eqnarray}%
\label{sp-eqns}%
\end{subequations}%
The parameter space of a triangular-wedge in this case is
$\tilde{s}\geq 0$, and 
$\tilde{p}\geq 2\tilde s + 2\sqrt{1+\tilde s^2} \geq\text{Max}(2,4\tilde s)$.
See FIG.~\ref{3dspTri-wed}. The inequality, $\tilde p > 4\tilde s$,
is a consequence of the triangle inequality.
\begin{center}
\begin{figure}
\includegraphics{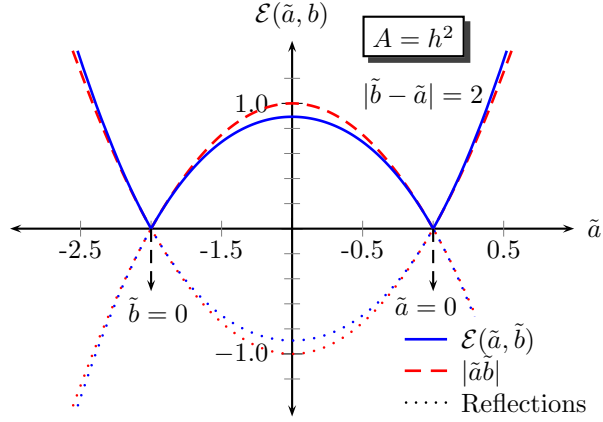}
\caption{
${\cal E}(\tilde a,\tilde b)$ as a function of  $\tilde a$ for fixed area,
$A=h^2$. The irreducible three-body Casimir energy is minimal when
the shorter side of the wedge is perpendicular to the Dirichlet plate
($\tilde a=0$ or $\tilde b=0$).
The maximum in the intermediate region corresponds to the unstable
equilibrium of an isosceles triangle. The dashed curves are the 
approximation ${\cal E}(\tilde a,\tilde b)\sim |\tilde a\tilde b|$
obtained by replacing the integrals in Eq.~(\ref{E123=tri-wed}) with unity.
The dotted curves are reflections about the ${\cal E}=0$ line.}
\label{area-fix-fig}
\end{figure}
\end{center}
In FIG.~\ref{area-fix-fig} we plot the energy as a function of
$\tilde a$ for fixed
area: $A=h^2$, or $|\tilde b - \tilde a|=2$, or $|\tan\beta-\tan\alpha|=2$.
The three-body Casimir energy for a waveguide of given cross-sectional
area is minimal
when the shorter side of the wedge is perpendicular to the Dirichlet plate
($\alpha=0$ or $\beta=0~[\alpha=\tan^{-1}(-2)]$). 
In the intermediate region the energy is extremal for an isosceles
triangle $(-\tilde a =\tilde b=1$) with ${\cal E}(-1,1)=0.893112\ldots$.
The dashed curve in FIG.~\ref{area-fix-fig} represents the
approximation ${\cal E}(\tilde a, \tilde b)\sim |\tilde a\tilde b|$
obtained by setting the dimensionless integral in Eq.~(\ref{E123=tri-wed})
to 1. Remarkably, this extremely simple expression for
the irreducible three-body energy is accurate to better than ten percent
everywhere. We also show reflections of the curves to illustrate that the
discontinuities in the slope are entirely due to the absolute value
in the pre-factor $|\tilde a \tilde b|$ and the integral itself is analytic. 

We rewrite the irreducible three-body Casimir energy as a function 
of the cross-sectional area and perimeter by inverting
Eqs.~(\ref{sp-eqns}) to obtain
\begin{subequations}
\begin{eqnarray}
\tilde{a} &=& \begin{cases}
\pm\tilde{\mu}-\tilde{s},\qquad\text{if}\quad \tilde{b}>\tilde{a}, \\ 
\pm\tilde{\mu}+\tilde{s},\qquad\text{if}\quad \tilde{b}<\tilde{a}, \\ 
\end{cases} \\
\tilde{b} &=& \begin{cases}
\pm\tilde{\mu}+\tilde{s},\qquad\text{if}\quad \tilde{b}>\tilde{a}, \\ 
\pm\tilde{\mu}-\tilde{s},\qquad\text{if}\quad \tilde{b}<\tilde{a}, \\
\end{cases}
\end{eqnarray}
\label{ab=spmu}
\end{subequations}
where
\begin{equation}
\tilde{\mu}= \frac{1}{2\tilde{p}}
\frac{(\tilde{p}-2\tilde{s})}{(\tilde{p}-4\tilde{s})}
\bigg[ \tilde{p} (\tilde{p}-4\tilde{s})
\Big\{ \tilde{p} (\tilde{p}-4\tilde{s}) -4\Big\}
\bigg]^\frac{1}{2}.
\end{equation}
Substituting Eqs.~(\ref{ab=spmu}) in Eq.~(\ref{E123=tri-wed}),
the three-body Casimir energy as of function of perimeter and area is
\begin{equation}
{\cal E}(\tilde s,\tilde p) =|\tilde{\mu}^2-\,\tilde{s}^2|
\int_0^1 \int_0^1 \frac{du_1 du_2}{\bar{u}_{12}^2}
\,Q\left( \frac{u_{12}^2}{\bar{u}_{12}^2} \right),
\label{E123sp=tri-wed}
\end{equation}
where the rescaled distances in terms of area and perimeter are given by
\begin{subequations}
\begin{eqnarray}
\bar{u}_{12}^2 &=& [\tilde{\mu} (u_1-u_2) +\tilde{s}(u_1+u_2) ]^2 
+ [|1-u_1| +|1-u_2|]^2,\\
u_{12}^2 &=& [\tilde{\mu} (u_1-u_2) +\tilde{s}(u_1+u_2) ]^2 + (u_1 -u_2)^2.
\end{eqnarray}
\end{subequations}
In FIG.~\ref{3dspTri-wed}  the irreducible three-body contribution
to the vacuum energy of a semiweak wedge is plotted
as a function of dimensionless area and perimeter of the cross-section.
The energy now is minimal along the curve 
\begin{equation}
\tilde{p}=1+2\tilde{s}+\sqrt{1+4\tilde{s}^2}
=\begin{cases}
2+2\tilde{s}+{\cal O}(2\tilde{s})^2 \qquad & 2\tilde{s}<1,\\
1+4\tilde{s}+{\cal O}\left(\frac{1}{2\tilde{s}}\right) \qquad & 2\tilde{s}>1,
\end{cases}
\end{equation}
which corresponds to right-angled triangles.
The energy diverges along the line $\tilde{p}\geq2$, $\tilde s=0$,
which corresponds to the two sides of the wedge coinciding
($\alpha=\beta$). In FIG.~\ref{3dspTri-wed} the curve 
$\tilde{p}=1+2\tilde{s}+\sqrt{1+4\tilde{s}^2}$ for $s\geq 0$
corresponds to right triangles of minimal energy, and the boundary of
the parameter space at $\tilde p =2\tilde{s}+2\sqrt{1+\tilde{s}^2}$
for $\tilde s \geq0$ is associated with isosceles triangles.

We do not observe that ${\cal E}(\tilde s,\tilde p)$ is a function of 
$\tilde p^2/\tilde s$ only.
The rather good approximation obtained by ignoring the
dependence on the integral in Eq.~(\ref{E123sp=tri-wed}),
suggests that the energy approximately is given by
\begin{equation}
{\cal E}(\tilde s,\tilde p) \sim |\tilde{\mu}^2 -\tilde{s}^2|
=\frac{ \big| 4(\tilde p-2\tilde s)^2 -\tilde p^2 (\tilde p-4\tilde s)^2 \big|
}{4\tilde p (\tilde p -4\tilde s)},
\end{equation}
a somewhat involved function of the perimeter and area.


\section{Parabolic-wedge on a Dirichlet plate}
\label{par-wedge:s}

\begin{center}
\begin{figure}
\includegraphics{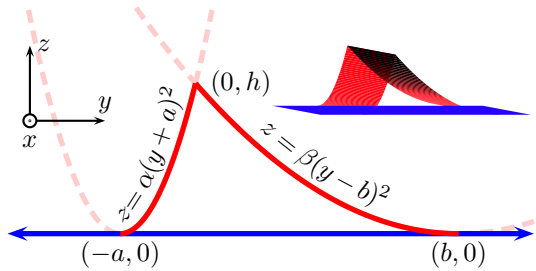}%
\caption{Weakly interacting parabolic-wedge on a Dirichlet plate.}
\label{parwedoTri}
\end{figure}
\end{center}

To explicitly verify that not just corner divergences have been subtracted
in the irreducible three-body contribution to the 
vacuum energy~\cite{Schaden:2010wv},
we also consider a weakly interacting parabolic-wedge 
on a Dirichlet plate. It is described by the potentials
\begin{subequations}
\begin{eqnarray}
V_1(y,z) &=& \lambda_1 \delta (-z+\alpha(y-a)^2) \,\theta_1, 
\quad \text{with} \quad 
\theta_1\equiv\theta(y-\text{min}[0,a])\,\theta(\text{max}[0,a]-y), \\
V_2(y,z) &=& \lambda_2 \delta (-z+\beta(y-b)^2) \,\theta_2, 
\quad \text{with} \quad 
\theta_2\equiv\theta(y-\text{min}[0,b])\,\theta(\text{max}[0,b]-y),\\
V_3(z) &=& \lambda_3 \delta (z), \quad \text{with} \quad \lambda_3\to\infty.
\end{eqnarray}
\label{parwed-pot-w}%
\end{subequations}%
The parameters $\alpha$ and $\beta$ here give the foci of the parabolas 
and have dimensions of inverse length. 
The constraint $\alpha a^2=\beta b^2=h$ implies that the two parabolas
intersect at $(y=0,z=h)$. See FIG.~\ref{parwedoTri}.
As in the case of the wave guide with triangular cross-section,
the base has length $|b-a|$ and the height of the wedge above the 
Dirichlet plate is $h$. The parameter regions are: 
$-\infty<a,b<\infty$, or, equivalently, $0\leq\alpha,\beta<\infty$.
We measure lengths in multiples of $h$ and
use parameters $(h,\tilde a=a/h,\tilde b=b/h)$ to describe it.

\begin{center}
\begin{figure}
\includegraphics{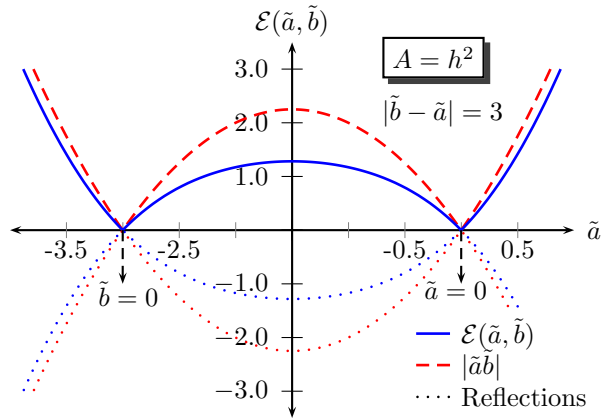}
\caption{Irreducible three-body Casimir energy for a parabolic waveguide
of given cross-sectional area. See FIG.~\ref{area-fix-fig} for description.}
\label{par-area-fix-fig}
\end{figure}
\end{center}

We proceed exactly as for the triangular-wedge and find that the 
three-body Casimir energy of a parabolic-wedge is also given by 
Eq.~(\ref{E123=tri-wed}), except that the distances now are given by 
\begin{subequations}
\begin{eqnarray}
\bar{u}_{12}^2 &=& (\tilde{a} u_1-\tilde{b} u_2 )^2+[(1-u_1)^2 +(1-u_2)^2]^2,\\
u_{12}^2 &=& (\tilde{a} u_1-\tilde{b} u_2)^2+[(1-u_1)^2 -(1-u_2)^2]^2.
\end{eqnarray}
\end{subequations}
The three-body Casimir energy of a parabolic-wedge on a Dirichlet plate also
is minimized when either $\tilde{a}=0, \alpha=\infty$, 
or $\tilde{b}=0, \beta=\infty$. Due to the constraint, 
$\alpha a^2=\beta b^2=h$, the shorter side of the parabolic wedge in this 
case degenerates to a straight line perpendicular to the Dirichlet plate. 
Most of the analysis of the waveguide with two sides of parabolic 
cross-section is the same as for a 
triangular one with only minor changes in interpretation.
We note that the rescaled area and perimeter of the parabolic wedge are 
\begin{subequations}
\begin{eqnarray}
\tilde{s}&=&\frac{1}{3}|\tilde{b}-\tilde{a}|,\label{par-s=ab} \\
\tilde{p} &=& |\tilde{b}-\tilde{a}| +\frac{\tilde{a}}{2} 
\bigg[ \frac{\tilde{a}^2}{2}\sinh^{-1}\frac{2}{\tilde{a}^2}
+ \sqrt{1+\frac{4}{\tilde{a}^4}} \bigg]
+ \frac{\tilde{b}}{2}
\bigg[ \frac{\tilde{b}^2}{2}\sinh^{-1}\frac{2}{\tilde{b}^2}
+ \sqrt{1+\frac{4}{\tilde{b}^4}} \bigg].%
\label{par-p=ab}
\end{eqnarray}%
\label{par-sp-eqns}%
\end{subequations}%
The three-body Casimir energy of a semiweak parabolic-wedge is
shown in FIG.~\ref{par-area-fix-fig}. 
The approximation of replacing the integral by unity
is not very accurate in this case, but the overall dependence of the 
irreducible three-body energy of a parabolic waveguide on the parameters
$\tilde a$ and $\tilde b$ is qualitatively similar to that of a
triangular one.

\section{Discussion}
In Casimir studies one generally is interested in dependence of the 
vacuum energy of massless quantum fields on the presence of objects
whose interaction with the quantum fields is treated semiclassically,
with quantum fluctuations of the fields describing the objects themselves
being disregarded. This leads to an effective action with ultraviolet 
divergent contributions associated with geometrical properties of the 
objects reflected by the coefficients~\cite{weyl,Minakshisundaram:1949xg}
in the asymptotic expansion of the heat 
kernel~\cite{Kac:1966xd,Fulling:2003zx,kirsten2002spectral}. 
The corresponding ultra-violet divergences in the vacuum energy are 
proportional to the spatial volume, surface areas, curvatures, 
as well as number and type of corners or intersections of the objects.
They depend only on \emph{local} geometric properties of the system.

Fortunately, there also are non-local contributions to the vacuum 
energy that do not depend on the high energy description of the 
model and can be reliably obtained in semiclassical 
approximation. The best known of these is the force between disjoint
classical objects due to vacuum fluctuations, first obtained 
for parallel metallic plates by Casimir~\cite{Casimir:1948dh}. It has 
since been shown that this force is always finite~\cite{Kenneth:2006vr}
and that the associated finite contribution to the vacuum energy may be 
computed for arbitrary objects in terms of the single-body scattering 
matrices with Eq.~(\ref{trloggen}). 
The investigation of generalized pistons 
in~\cite{Schaden:PhysRevLett.102.060402,Schaden:PhysRevA.79.052105} 
suggested that one may isolate finite parts of the vacuum energy
that describe the forces between objects even if these are not
mutually disjoint. These ideas were formalized in \cite{Schaden:2010wv}
where irreducible N-body contributions to the vacuum energy were
defined recursively and shown to be finite unless the N-bodies have 
a common intersection. For a scalar field whose interaction with N-objects
are semiclassically described by positive local potentials,
the irreducible contribution to the vacuum energy furthermore
was found to be positive for an odd, and negative for an even,
number of objects.

We have here put these general considerations on a much more
practical and concrete footing and developed a formalism to extract 
and compute irreducible $N$-body contributions from the single-body
transition matrices. Starting from Faddeev's equations in Eq.~(\ref{fadeqn}),
the irreducible parts of the $N$-body scattering matrix were extracted
recursively. We used this formalism to compute several examples
of irreducible two--and three-body Casimir energies. All our two-body 
results have been obtained previously, but we were able to reproduce
some of them in a much simpler and direct manner. 
Our three-body results for irreducible Casimir energies are new.
The irreducible three-body contributions
to the Casimir energy of parallel semitransparent plates was obtained
analytically and indeed remains finite when two of the three plates
overlap. We showed explicitly how the irreducible three-body contribution
precisely cancels the irreducible two-body Casimir energy of the outer
plates when Dirichlet boundary conditions are imposed on the central 
plate--providing a raison-d'{\^e}tre for both, the existence, and sign, 
of the three-body contribution to the force. For semitransparent plates
the cancellation is not complete but can be sizable.

In Section \ref{3body-sweak:s} we analyze
the irreducible three-body interaction in semiweak approximation.
In this approximation we are able to compute the irreducible 
three-body Casimir energy for objects that are not mutually disjoint
and whose irreducible two-body contributions diverge. The irreducible 
three-body contributions to the vacuum energy of a waveguide constructed by 
placing a weakly interacting triangular--or parabolic-wedge on top of 
a Dirichlet plate was found to be finite and computable without intermediate 
regularization. 
Our examples demonstrate that not only corner divergences, but also
divergences related to curvature are subtracted by this
procedure. We also explicitly verified that the irreducible
three-body contribution to the vacuum energy of a massless scalar
field is positive.

To develop a better understanding in a non-perturbative setting,
we are currently investigating the irreducible three-body vacuum energy
of a triangular waveguide formed by imposing Dirichlet boundary conditions
on three intersecting infinite planes (the geometry is similar
to that of FIG.~\ref{wedoTri}, but with sides of infinite extent). 
In the limit of an extremely flat triangular cross-section, we intend
to compare the numerical results with analytic calculations. We 
further wish to extend these methods to the physically relevant 
electromagnetic case. Although irreducible three-body contributions
to the vacuum energy are expected to remain finite, we so far have
no rigorous statements about their sign for vector fields. Interestingly,
it is at least conceptually feasible to directly measure irreducible
electromagnetic three-body contributions to the vacuum energy by 
balancing off irreducible two-body parts. We are investigating whether 
this is experimentally feasible.


\begin{acknowledgments}
KVS would like to thank Prachi Parashar for useful comments and suggestions
at various stages of this project.
This work was supported by the National Science Foundation with 
Grant no.~PHY0555580.
\end{acknowledgments}

\bibliography{%
              biblio/b7001-many-body,%
              biblio/b7002-Martins-data%
              }

\end{document}